\documentclass{llncs}
\usepackage{pstricks}
\usepackage{colortbl}

\usepackage{pst-node}
\usepackage{pst-grad}
\usepackage{pst-coil}
\usepackage{pst-text}
\usepackage{pst-rel-points}
\usepackage{amssymb}
\usepackage{amsmath}
\usepackage{times}
\usepackage{calc}
\usepackage{etex}
\usepackage{ifthen}
\usepackage{multirow}
\usepackage{rotating}
\usepackage{longtable}
\usepackage[normalem]{ulem}
\usepackage{stmaryrd}
\usepackage{xspace}
\usepackage{hyperref}

\newcommand{\callpt}[1]{\mbox{$c_{#1}$}}
\newcommand{\retpt}[1]{\mbox{$r_{#1}$}}
\newcommand{\End}[1]{\text{{\sf\em end}$_{#1}$}\xspace}
\newcommand{\Start}[1]{\text{{\sf\em start}$_{#1}$}\xspace}
\newcommand{\boundary}{\text{\sf\em BI}\xspace}
\newcommand{\pred}{\text {\sf\em{pred}\/}}
\newcommand{\Succ}{\text{\sf\em {succ}\/}}
\newcommand{\In}[1]{\text{\sf\em In\/$_{#1}$}}
\newcommand{\Out}[1]{\text{\sf\em Out\/$_{#1}$}}

\newcommand{\var}{\text{\sf\bfseries\em V}\xspace}
\newcommand{\pointer}{\text{\sf\bfseries\em P}\xspace}

\psset{unit=1mm}

\newcommand{\pt}[2]{\text{$(#1,#2)$}}

\newcommand{\aptE}{\text{$\mathcal{A}$}\xspace}

\newcommand{\lvE}{\text{$\mathcal{L}$}\xspace}

\newcommand{\rep}{\text{{\sf\em Rep\/}}\xspace}
\newcommand{\reg}{\text{{\sf\em Reg\/}}\xspace}
\newcommand{\lin}{\text{{\sf\em Lin\/}}\xspace}
\newcommand{\ain}{\text{{\sf\em Ain\/}}\xspace}
\newcommand{\uin}{\text{{\sf\em Uin\/}}\xspace}
\newcommand{\lout}{\text{{\sf\em Lout\/}}\xspace}
\newcommand{\aout}{\text{{\sf\em Aout\/}}\xspace}
\newcommand{\uout}{\text{{\sf\em Uout\/}}\xspace}

\newcommand{\must}{\text{\sf\em Must\/}\xspace}
\newcommand{\Def}{\text{{\sf\em Def}}\xspace}
\newcommand{\Pointee}{\text{{\sf\em Pointee}}\xspace}
\newcommand{\Kill}{\text{{\sf\em Kill}}\xspace}
\newcommand{\Ref}{\text{{\sf\em Ref}}\xspace} 

\newcommand{\lrestrict}[2]{\text{$\protect#1{\mid}_{\protect#2}$}}
\newcommand{\bigrestrict}[2]{\text{$\protect#1\!\mbox{\Large$\mid_{\protect#2}$}$}}

\protect{\newlength{\BRACEL}}
\newcommand{\OB}{\{\hspace*{\BRACEL}}
\newcommand{\CB}{\}}
\protect{\newlength{\TAL}} 
\newcommand{\NL}[1]{\hspace*{#1\TAL}}
\newlength{\codeLineLength}
\newcommand{\codeLine}[3]{%
\psframebox[framesep=0,fillstyle=solid,fillcolor=#3,
	linestyle=none]{\makebox[\codeLineLength][l]{%
	\rule[-.3em]{0em}{1.em}{\em%
	 \NL{#1}{\mbox{#2}}}}}
\\ }

\title{Lazy Pointer Analysis}
\author{Uday P. Khedker\inst{1} \and Alan Mycroft\inst{2} \and Prashant Singh Rawat\inst{1}\thanks{Supported 
by GCC Resource Center, funded by Dept. of Information  Technology, Gov. of India
as a part of the
National Resource Center for Free and Open Source Software
(NRCFOSS).}
}
\institute{Indian Institute of Technology Bombay \\
\email{\{uday,prashantr\}@cse.iitb.ac.in}
\and University of Cambridge \\
\email{Alan.Mycroft@cl.cam.ac.uk}
}

\newcommand{\dfv}[2]{\text{$(\protect#1,\{\protect#2\})$}}
\newcommand{\sdfv}[2]{\text{$(\protect#1,\protect#2)$}}
\newcommand{\dfvl}[2]{\text{$(\protect#1,\protect#2)_L$}}
\newcommand{\dfva}[2]{\text{$(\protect#1,\protect#2)\!_A$}}

\begin{document}

\maketitle

\begin{abstract}
Flow- and context-sensitive pointer analysis is generally considered
too expensive for large programs; most tools relax one or both of the
requirements for scalability. We formulate a flow- and context-sensitive
points-to analysis that is lazy in the following sense: points-to
information is computed only for live pointers and its propagation is
sparse (restricted to live ranges of respective pointers). Our analysis
also: ({\em i\/}) uses strong liveness, effectively including dead
code elimination; ({\em ii\/}) afterwards calculates must-points-to
information from may-points-to information instead of using a mutual
fixed-point; ({\em iii\/}) uses value-based termination of call strings
during interprocedural analysis (which reduces the number of call
strings significantly).

A naive implementation of our analysis within GCC-4.6.0 gave analysis
time and size of points-to measurements for SPEC2006. Using liveness
reduced the amount of points-to information by an order of magnitude
with no loss of precision. For all programs under 30kLoC we found that
the results were much more precise than gcc's analysis. What comes as
a pleasant surprise however, is the fact that below this cross-over
point, our naive linked-list implementation is faster than a flow-
and context-insensitive analysis which is primarily used for efficiency.
We speculate that lazy flow- and context-sensitive analyses may be not
only more precise, but also more efficient, than current approaches.
\end{abstract}

\section{Introduction}
\label{sec:intro}

Interprocedural data flow analysis extends the scope of analysis across
procedure boundaries to incorporate the effect of callers 
on callees and vice-versa. The efficiency and scalability of such
an analysis is a major concern. The precision of such an analysis
requires flow-sensitivity (associating different
information with distinct control flow points) and context-sensitivity
(computing information depending upon the calling context). Sacrificing
precision for scalability is a common trend in interprocedural data
flow analysis. This is more prominent in pointer analysis in which the
size of information could be large. Flow- and context-sensitive pointer
analysis is considered prohibitively expensive and most methods relax
one or both of the requirements for scalability.

We formulate a flow- and context-sensitive points-to analysis that is
lazy: points-to information is computed only for the pointers that are
live and the propagation of points-to information is sparse in that
it is restricted to live ranges of respective pointers. We use strong
liveness which identifies the pointers that are directly used or are
used in defining pointers that are strongly live. Thus strong liveness
incorporates the effect of dead code elimination on liveness and is more
precise than simple liveness.

\begin{figure}[t]
\begin{tabular}{@{}c|c|@{\ \ }c}
\setlength{\codeLineLength}{23mm}
\begin{tabular}{@{}c}
	\codeLine{1}{main()}{white}
	\codeLine{1}{\OB \,x = \&y;}{white}
	\codeLine{2}{w = \&x;}{white}
	\codeLine{2}{p();}{white}
	\codeLine{2}{print z;}{white}
	\codeLine{1}{\CB}{white}
	\codeLine{1}{}{white}
	\codeLine{1}{p()}{white}
	\codeLine{1}{\OB if (...)}{white}
	\codeLine{2}{\OB z = w;}{white}
	\codeLine{3}{p();}{white}
	\codeLine{3}{z = $*$z;}{white}
	\codeLine{2}{\CB}{white}
	\codeLine{1}{\CB}{white}
	\end{tabular}
&
\begin{tabular}{@{}c@{}}
\begin{pspicture}(0,0)(48,56)
\putnode{entry}{origin}{8}{53}{\psframebox{\Start{m}}}
\putnode{w}{entry}{7}{0}{$\;1$}
\putnode{n1}{entry}{0}{-8}{\psframebox{$x = \&y$}}
\putnode{w}{n1}{8}{0}{$\;2$}
\putnode{n2}{n1}{0}{-8}{\psframebox{$w = \&x$}}
\putnode{w}{n2}{8}{0}{$\;3$}

\putnode{c1}{n2}{0}{-8}{\psframebox{$\;\;\;c_1\;\;\;$}}
\putnode{w}{c1}{7}{0}{$\;4$}
\putnode{r1}{c1}{0}{-12}{\psframebox{$\;\;\;r_1\;\;\;$}}
\putnode{w}{r1}{7}{0}{$\;5$}
\putnode{n4}{r1}{0}{-8}{\psframebox{print {$z$}}}
\putnode{w}{n4}{7}{0}{$\;6$}
\putnode{exit}{n4}{0}{-8}{\psframebox{\End{m}}}
\putnode{w}{exit}{7}{0}{$\;7$}
\putnode{sp}{entry}{24}{-6}{\psframebox{\Start{p}}}
\putnode{w}{sp}{7}{0}{$\;8$}
\putnode{n5}{sp}{6}{-8}{\psframebox{$z=w$}}
\putnode{w}{n5}{-7}{0}{$9\;$}
\putnode{c2}{n5}{0}{-8}{\psframebox{$\;\;\;c_2\;\;\;$}}
\putnode{w}{c2}{-8}{0}{$10$}
\putnode{r2}{c2}{0}{-11}{\psframebox{$\;\;\;r_2\;\;\;$}}
\putnode{w}{r2}{-8}{0}{$11$}
\putnode{n6}{r2}{0}{-8}{\psframebox{$z = *z$}}
\putnode{w}{n6}{-8}{0}{$12\;$}
\putnode{ep}{n6}{-6}{-8}{\psframebox{\End{p}}}
\putnode{w}{ep}{7}{0}{$13$}
\ncline{->}{entry}{n1}
\ncline{->}{n1}{n2}
\ncline{->}{n2}{n3}
\ncline{->}{n2}{c1}
\ncline{->}{r1}{n4}
\ncline{->}{n4}{exit}
\ncline{->}{sp}{n5}
\ncline{->}{n5}{c2}
\ncline{->}{r2}{n6}
\ncline{->}{n6}{ep}
\nccurve[angleA=240,angleB=120,ncurv=.8,offset=-1,nodesep=.5]{->}{sp}{ep}
\psset{linestyle=dashed,dash=.8 .8}
\ncloop[angleA=270,angleB=90,loopsize=-8,linearc=.5,offsetB=1,arm=2.5]{->}{c2}{sp}
\ncloop[angleA=270,angleB=90,loopsize=-13,linearc=.5,offsetA=1,arm=2.5]{->}{ep}{r2}
\ncloop[angleA=270,angleB=90,loopsize=-12,linearc=.5,offsetB=-1,arm=2.5]{->}{c1}{sp}
\ncloop[angleA=270,angleB=90,loopsize=11,linearc=.5,offsetA=-1,arm=2.5]{->}{ep}{r1}
\end{pspicture}
\end{tabular}
&
\begin{minipage}{45mm}
Let \pt{a}{b} at a program point denote that $a$ points-to $b$ at that program point. Then,
\begin{itemize}
\item $z$ is live at the exit of $9$ which makes $w$ live at the exit of $3$.
      Hence we should compute \pt{w}{x} in $3$ and thereby \pt{z}{x} in $9$. 
      This causes $x$ to be live because of $*z$ in $12$. Hence we should compute 
      \pt{x}{y} in $2$ and \pt{z}{y} in $12$.
\item \pt{w}{x} and \pt{x}{y} should not be propagated to $5$, $6$, $7$ because $w,x$ are not live 
      in these nodes.
\end{itemize}
\end{minipage}
\end{tabular}
\caption{A motivating example for lazy points-to analysis, its supergraph representation and some observations.
      The solid edges in the supergraph represent intraprocedural control flow while dashed edges represent
      interprocedural control flow.}
\label{fig:ipa.pta}
\end{figure}
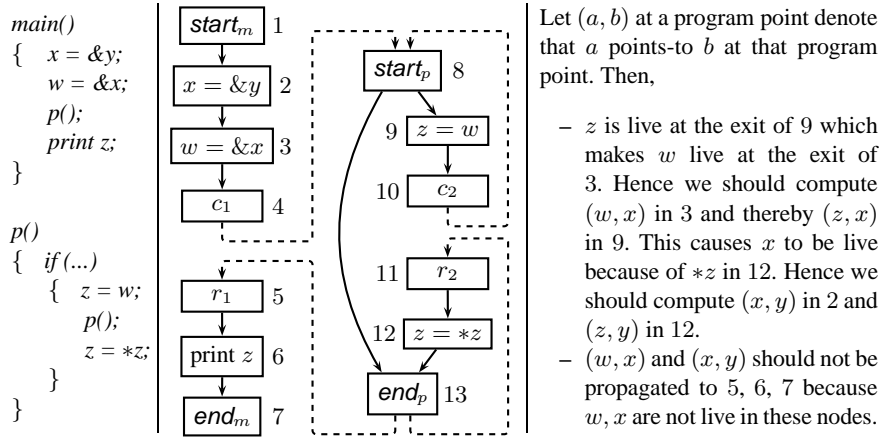

Fig.~\ref{fig:ipa.pta} provides a motivating example for lazy pointer
analysis.
By printing $z$, the main procedure makes $z$ live at node $12$
after the call to procedure $p$.
This makes $z$ in node $12$ live. This in turn
makes $w$ live in node $9$ and then in $3$ resulting in
the points-to pair \pt{w}{x}. This pair is propagated to $12$
giving the pair \pt{z}{x}. When this information becomes available
in $12$, $x$ becomes live. This liveness is propagated to $2$
giving the pair \pt{x}{y}. Eventually we get the pair \pt{z}{y}
in $12$.
Figures~\ref{fig:ipa.pta.result.1} and~\ref{fig:ipa.pta.result.2} give
fuller detail after formulating lazy pointer analysis interprocedurally.
Here we observe the following:
\begin{itemize}
\item {\em Lazy computation}. Points-to pairs are computed when the
      pointers become live.
\item {\em Sparse propagation}. Pairs \pt{x}{y} and \pt{w}{x} are not 
      propagated beyond the call to $p$ in the main procedure in spite of 
      the fact that $x$ or $w$ are not modified in $p$.
\item {\em Flow sensitivity}. Points-to information is different for
      different control flow points.
\item {\em Context sensitivity}. \pt{z}{x} holds only for the inner call to $p$
      made from within $p$ but not for the outer call to $p$ made from within the
      main procedure. Thus in spite of $z$ being live in $6$, 
      \pt{z}{x} is not propagated to $6$ but \pt{z}{y} is.
\end{itemize}

We propose a novel data flow framework that employs an
interdependent formulation for discovering strong liveness and
points-to information for pointer variables. This framework computes
must-points-to information from may-points-to information without
requiring an additional fixed-point computation. At the interprocedural
level, flow- and context-sensitivity is ensured by using value-based
termination of call strings. 

Our findings conclusively demonstrate that instead of achieving scalability
by compromising on precision, it is far better to contain the explosion
of information by clearly distinguishing between the information that
is relevant from the information that is not relevant. Since pointer
information is required to uncover the data items that are accessed
indirectly or functions that are invoked indirectly, it is relevant
only when there is some use of a pointer. We show that this change in
perspective provides significant benefits in terms of time and space
requirements of pointer analysis.

The rest of the paper is organised as follows:
Section~\ref{sec:background} reviews the background. 
Section~\ref{sec:lazy.pta} formulates the mutually dependent
liveness and points-to analysis at the
intraprocedural level. 
Section~\ref{sec:properties}
formalises and proves some important properties of our analysis.
It is lifted to interprocedural level
in Section~\ref{sec:interprocedural}. 
Section~\ref{sec:related.work} discusses the related work while
Section~\ref{sec:measurements} presents the empirical data.
 Section~\ref{sec:conclusions} concludes the
paper.

\section{Background}
\label{sec:background}

This section reviews intra- and interprocedural data flow analysis
and pointer analysis.

\newcommand{\LD}{\text{{\bfseries\em L}$_G$}\xspace}
\newcommand{\FD}{\text{{\bfseries\em F}$\!_G$}\xspace}

\subsubsection{Overview of Data Flow Analysis.}
\label{sec:dff}
Data flow analysis is formulated in terms of {\em data flow equations\/} that
describe how the required {\em data flow information\/} can be computed for a statement.
The set of data flow values, the functions to compute them and the operation
to merge them are described by a {\em data flow framework\/}.

Unlike the classical view that treats a data flow framework and its
instance as distinct\cite{%
Kildall.GA:1973:unified-approach-to,%
Kam.JB.Ullman.JD:1977:Monotone-data-flow,%
Hecht.MS:1977:Flow-Analysis-of,%
Aho.AV.Lam.MS.Sethi.R.Ullman.JD:2006:Compilers-Principles-Techniques,%
Nielson.F.Nielson.HR.Hankin.C:1998:Principles-of-Program},
we view a data flow framework parameterised by a program because the
data flow values and the functions that manipulate them depend on the program being 
analysed. Formally, a data flow framework is a tuple 
\text{$\langle\LD,\sqcap_G,\FD\rangle$}~\cite{%
Khedker.U.Sanyal.A.Karkare.B:2009:Data-Flow-Analysis} 
where $G$ is an unspecified graph representing a program,
\LD is a meet semilattice representing the data flow values relevant to the analysis,
and \FD is a set of admissible flow functions from \LD to \LD. 
We require all strictly descending chains in \LD to be finite.
$\sqcap_G$ is the meet operator of \LD.
We require the flow functions in \FD to be monotonic.

At the intraprocedural level, a procedure is represented by a {\em
control flow graph\/} (CFG) whose nodes represent program statements
and edges represent control transfers. A CFG for procedure $p$ must
satisfy the following requirements: there must be a unique entry node
\Start{p} with no predecessor and a unique exit node \End{p} with no
successor, each node $n$ must be reachable from \Start{p}, and \End{p}
should be reachable from each node. At the interprocedural level, a
program is represented by a {\em supergraph\/} which connects the CFGs
by interprocedural edges. A call to procedure $p$ at call site $i$ is
split into a {\em call node\/} \callpt{i} and a {\em return node\/}
\retpt{i} with a call edge \text{$ \callpt{i} \rightarrow \Start{p}$}
and a return edge \text{$\End{p} \rightarrow \retpt{i}$}. In examples
we number nodes in the CFG in reverse post-order and assign contiguous
numbers across procedures.
Fig.~\ref{fig:ipa.pta}
provides an example of a supergraph.

\begin{figure}[t]
\centering
\begin{tabular}{c|c}
Forward Analysis (\In{n} influences \Out{n})
	\;\;&\;\;
Backward Analysis (\Out{n} influences \In{n})
\\\hline
$
\begin{array}{rcl}
\In{n} & = 
	&
	\left\{
	\begin{array}{c@{ \ \ \ \ }l}
	\boundary & n = \Start{p}
		\\
	\displaystyle\bigsqcap_{p \in \pred(n)} \Out{p}	
		& \text{otherwise}
	\end{array}
	\right.
	\\
\Out{n} & = 
	&
	f_n (\In{n})
\end{array}
$
&
$
\begin{array}{rcl}
\In{n} & = 
	&
	f_n (\Out{n})
	\\
\Out{n} & = 
	&
	\left\{
	\begin{array}{c@{ \ \ \ \ }l}
	\boundary & n = \End{p}
		\\
	\displaystyle\bigsqcap_{s \in \Succ(n)} \In{s}	
		& \text{otherwise}
	\end{array}
	\right.
\end{array}
$
\end{tabular}
\caption{Typical data flow equations for some procedure $p$.}
\label{fig:dfeq}
\end{figure}

Data flow equations (Fig.~\ref{fig:dfeq}) define data flow variables
\In{n} and \Out{n} which represent the data flow information associated
with the entry and exit points of node $n$. \text{$\In{n}, \Out{n}
\in \LD$} and \text{$f_n \in \FD$}. The {\em boundary information\/}
\boundary represents the data flow information at the procedure entry
for forward analysis and procedure exit for backward analysis. Its value
is governed by the semantics of the information being discovered.
Interprocedural analysis eliminates the need for a fixed \boundary
(except for arguments to {\tt main}) and computes it from the calling
contexts during the analysis.

{\em Iterative methods\/} solve the data flow
equations by refining the values starting from a conservative
initialisation of $\top$. {\em Round robin\/} methods traverse the CFG
in a fixed order;
{\em work list\/} methods
maintain a list of the nodes whose values are to be recomputed.

\subsubsection{Interprocedural Data Flow Analysis.}
\label{sec:ipa.callstrings}
A supergraph contains control
flow paths which violate nestings of matching call return pairs (e.g.\
1-2-3-4-8-13-11 for the supergraph in Fig.~\ref{fig:ipa.pta}). Such
paths correspond to infeasible contexts. An interprocedurally valid path
is a feasible execution path containing a legal sequence of call and
return edges.

A {\em context-sensitive\/} analysis retains sufficient
information about calling contexts to distinguish the data flow
information reaching a procedure along different call chains. This
restricts the analysis to interprocedurally valid paths and 
ensures propagation of information from a callee to
appropriate call sites.
A {\em context-insensitive\/} analysis
does not distinguish between valid and invalid paths 
effectively merging data flow
information across calling contexts. Although the resulting information
is provably safe, it is often imprecise. Recursive procedures have
potentially infinite contexts, yet context-sensitive analysis is decidable
for data flow frameworks with finite lattices and it is sufficient to
maintain a finite number of contexts for such frameworks. However, this
number is combinatorially large even for non-recursive programs.
{\em Flow-insensitive\/} approaches disregard intraprocedural control flow
for efficiency. Instead of information being associated with each
program point, a single {\em summary} is computed. Although the summary
information is provably safe, it is imprecise. A {\em flow-sensitive\/} analysis
honours the control flow and computes data flow information separately
for each program point.


We use a flow- and context-sensitive 
approach called the {\em call-strings method\/}~\cite{%
Sharir.M.Pnueli.A:1981:Two-Approaches-to,%
Khedker.U.Sanyal.A.Karkare.B:2009:Data-Flow-Analysis,%
Khedker.UP.Karkare.B:2008:Efficiency-Precision-Simplicity}. It embeds
context information in the data flow information and ensures the
validity of interprocedural paths by maintaining a history of calls in
terms of call strings. A {\em call string\/} at node $n$ is a sequence
$c_1c_2\ldots c_k$ of call sites corresponding to unfinished calls
at $n$ and can be viewed as a snapshot of the call stack. $\lambda$
denotes an empty call string. Some call strings for our supergraph
in Fig.~\ref{fig:ipa.pta} are: $\lambda$, $c_1$, \text{$c_1c_2$},
\text{$c_1c_2c_2$} etc.

Call string construction is governed by interprocedural edges. 
Let $\sigma$ be a call string reaching procedure $p$. For
an intraprocedural edge $m\rightarrow n$ in $p$, $\sigma$ reaches $n$
unmodified. For a call edge \text{$\callpt{i}\rightarrow\Start{q}$}
where \callpt{i} belongs to $p$, call string $\sigma c_i$ reaches
\Start{q}. For a return edge \text{$\End{p}\rightarrow \retpt{j}$} where
\retpt{j} belongs to a caller of $p$, if the last call site in $\sigma$
is $c_j$ then the longest prefix that excludes $c_j$ reaches the call
site corresponding to \retpt{j}. If the last call site in $\sigma$ is
not $c_j$, the call string and its associated data flow value is not
propagated to the call site corresponding to \retpt{j}. This ensures
that the data flow information is only propagated to appropriate call sites.
In a backward analysis, the call string grows on traversing a return
edge and shrinks on traversing a call edge.

The augmented data flow information is a pair \text{$\langle \sigma,
d \rangle$} where $d$ is the data flow value propagated along call
string $\sigma$ and is modified by an intraprocedural edge only. A
work-list-based iterative algorithm is used to perform the data flow
analysis. The process terminates when no new pair \text{$\langle \sigma,
d \rangle$} is computed; merging the data flow values associated with
all call strings reaching node $n$ gives the final data flow value at
$n$. This method computes a safe and precise solution because it matches
call and return nodes in a path thereby excluding interprocedurally
invalid paths and traversing valid paths only.

In non-recursive programs, since the call strings are acyclic (no call
site occurs multiple times),
their number is finite and all of them are generated during analysis.
However, in recursive programs, new call strings are generated with every
visit to a call node involved in recursion. In such cases, the
number of call strings considered must be bounded using explicit criteria.
For computing a safe and precise solution, the full call-strings
method~\cite{Sharir.M.Pnueli.A:1981:Two-Approaches-to}
requires construction of all call strings of length up to $K \times
(|L|+1)^2$ where $K$ is the maximum number of distinct call sites in any
call chain and $L$ is the lattice of data flow values. For bit-vector
frameworks, we need to consider only those call strings in which a call
site appears at most
thrice~\cite{Karkare.B.Khedker.UP:2007:Improved-Bound-for}.
Since these numbers are very large for practical programs and we use
a recent variant in which the termination of call-string construction is based
on the equivalence of data flow values instead of precomputed length
bounds~\cite{Khedker.UP.Karkare.B:2008:Efficiency-Precision-Simplicity,%
Khedker.U.Sanyal.A.Karkare.B:2009:Data-Flow-Analysis}.
This allows us to discard call strings where they are redundant, and
regenerate them when required. For cyclic call strings representing
paths in recursion,
regeneration facilitates computation of data flow values
without explicitly constructing most of the call strings. This reduces
the space and time requirements of the analysis dramatically without
compromising on safety or precision.

\begin{figure}[t]
\begin{tabular}{@{}cc@{}}
\begin{tabular}{@{}c}
\begin{pspicture}(0,0)(29,48)
\putnode{n1}{origin}{13}{45}{\psframebox{$q = \&r$}}
\putnode{w}{n1}{8}{0}{1}
\putnode{n2}{n1}{0}{-9}{\psframebox{$p=q$}}
\putnode{w}{n2}{7}{0}{2}
\putnode{n3}{n2}{-7}{-8}{\psframebox{$p=*p$}}
\putnode{w}{n3}{7}{0}{\;3}
\putnode{n4}{n3}{0}{-8}{\psframebox{print $p$}}
\putnode{w}{n4}{7}{0}{4}
\putnode{n5}{n2}{7}{-12}{\psframebox{$s=q$\rule{0em}{.5em}}}
\putnode{w}{n5}{2}{5}{5}
\putnode{n6}{n4}{7}{-8}{\psframebox{$r=\&s$\rule{0em}{.3em}}}
\putnode{w}{n6}{8}{0}{6}
\putnode{n7}{n6}{0}{-9}{\psframebox{\white$p=*p$}\;}
\putnode{w}{n7}{8}{0}{7}
\ncline{->}{n1}{n2}
\ncline{->}{n2}{n3}
\ncline{->}{n2}{n5}
\ncline{->}{n3}{n4}
\ncline{->}{n4}{n6}
\ncline{->}{n5}{n6}
\ncline{->}{n6}{n7}
\ncloop[angleA=270,angleB=90,offset=2,loopsize=-13,arm=2.5,linearc=.5]{->}{n6}{n2}
\psset{framesep=.5,linestyle=none,fillstyle=solid,fillcolor=lightgray}
\end{pspicture}
\end{tabular}
&
$
\begin{array}{|c|c|c|c|c|}
\hline
\multirow{2}{*}{Node} 
	& \multicolumn{2}{c|}{\text{May analysis}}
	& \multicolumn{2}{c|}{\text{Must analysis}}
	\\ \cline{2-5}
	& \In{n}	& \Out{n}
	& \In{n}	& \Out{n}
	\\ \hline\hline
1 
	& \emptyset
	& \pt{q}{r}
	& \emptyset
	& \pt{q}{r}
	\\ \hline
2 
	& \renewcommand{\arraystretch}{.8}%
	  \begin{array}{c}
          \pt{p}{r},  \pt{p}{s},   \\
          \pt{q}{r}, 
          \pt{r}{s},  \pt{s}{r}
          \end{array}
	& \renewcommand{\arraystretch}{.8}%
	  \begin{array}{c}
          \pt{p}{r}, \pt{q}{r},  \\
	  \pt{r}{s}, \pt{s}{r}
          \end{array}
	& \renewcommand{\arraystretch}{.8}%
	  \begin{array}{c}
           \pt{q}{r}
          \end{array}
	& \renewcommand{\arraystretch}{.8}%
	  \begin{array}{c}
          \pt{p}{r}, \pt{q}{r}
          \end{array}
	\\ \hline
3 
	& \renewcommand{\arraystretch}{.8}%
	  \begin{array}{c}
          \pt{p}{r},  \pt{q}{r}, \\ \pt{r}{s},
          \pt{s}{r}
          \end{array}
	& \renewcommand{\arraystretch}{.8}%
	  \begin{array}{c}
          \pt{p}{s}, \pt{q}{r}, \\ \pt{r}{s},
	  \pt{s}{r}
          \end{array}
	& \renewcommand{\arraystretch}{.8}%
	  \begin{array}{c}
          \pt{p}{r}, \\ \pt{q}{r}
          \end{array}
	& \renewcommand{\arraystretch}{.8}%
	  \begin{array}{c}
          \pt{p}{s}, \pt{q}{r}
          \end{array}
	\\ \hline
4 
	& \renewcommand{\arraystretch}{.8}%
	  \begin{array}{c}
          \pt{p}{s},  \pt{q}{r}, \\ \pt{r}{s},
          \pt{s}{r}
          \end{array}
	& \renewcommand{\arraystretch}{.8}%
	  \begin{array}{c}
          \pt{p}{s}, \pt{q}{r}, \\ \pt{r}{s},
	  \pt{s}{r}
          \end{array}
	& \renewcommand{\arraystretch}{.8}%
	  \begin{array}{c}
          \pt{p}{s},  \\\pt{q}{r}
          \end{array}
	& \renewcommand{\arraystretch}{.8}%
	  \begin{array}{c}
          \pt{p}{s}, \pt{q}{r}
          \end{array}
	\\ \hline
5 
	& \renewcommand{\arraystretch}{.8}%
	  \begin{array}{c}
          \pt{p}{r},  \pt{q}{r}, \\ \pt{r}{s},
          \pt{s}{r}
          \end{array}
	& \renewcommand{\arraystretch}{.8}%
	  \begin{array}{c}
          \pt{p}{r}, \pt{q}{r}, \\ \pt{r}{s} ,
	  \pt{s}{r}
          \end{array}
	& \renewcommand{\arraystretch}{.8}%
	  \begin{array}{c}
          \pt{p}{r}, \\ \pt{q}{r}
          \end{array}
	& \renewcommand{\arraystretch}{.8}%
	  \begin{array}{c}
          \pt{p}{r}, \pt{q}{r}, \\ 
	  \pt{s}{r}
          \end{array}
	\\ \hline
6 
	& \renewcommand{\arraystretch}{.8}%
	  \begin{array}{c}
          \pt{p}{r},  \pt{p}{s}, \\ \pt{q}{r}, 
	  \pt{r}{s},  \pt{s}{r}
          \end{array}
	& \renewcommand{\arraystretch}{.8}%
	  \begin{array}{c}
          \pt{p}{r},  \pt{p}{s}, \\ \pt{q}{r}, 
	  \pt{r}{s},  \pt{s}{r}
          \end{array}
	& \renewcommand{\arraystretch}{.8}%
	  \begin{array}{c}
           \pt{q}{r}, 
          \end{array}
	& \renewcommand{\arraystretch}{.8}%
	  \begin{array}{c}
           \pt{q}{r}, 
	  \pt{r}{s}
          \end{array}
	\\ \hline
7 
	& \renewcommand{\arraystretch}{.8}%
	  \begin{array}{c}
          \pt{p}{r},  \pt{p}{s}, \\ \pt{q}{r}, 
	  \pt{r}{s},  \pt{s}{r}
          \end{array}
	& \renewcommand{\arraystretch}{.8}%
	  \begin{array}{c}
          \pt{p}{r},  \pt{p}{s}, \\ \pt{q}{r}, 
	  \pt{r}{s},  \pt{s}{r}
          \end{array}
	& \renewcommand{\arraystretch}{.8}%
	  \begin{array}{c}
          \pt{q}{r},\\ 
	  \pt{r}{s}
          \end{array}
	& \renewcommand{\arraystretch}{.8}%
	  \begin{array}{c}
          \pt{q}{r}, 
	  \pt{r}{s}
          \end{array}
	\\ \hline
\end{array}
$
\end{tabular}

\caption{An example of flow-sensitive intraprocedural points-to analysis.}
\label{fig:intra.pta.conventional}
\end{figure}
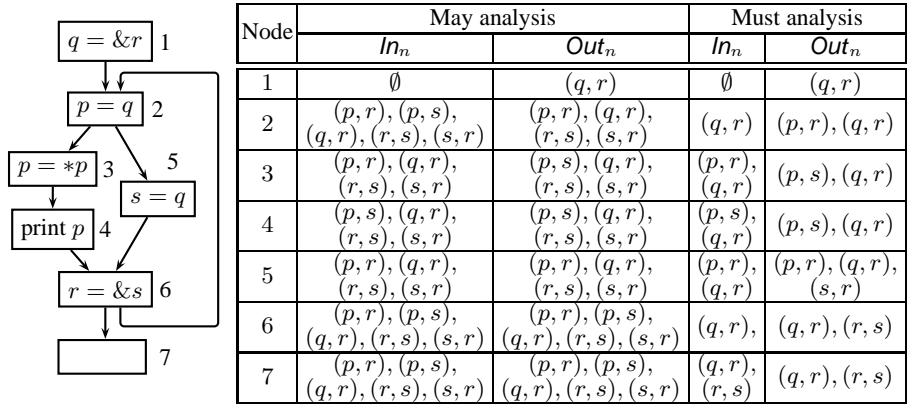

\subsubsection{Pointer Analysis}

Two forms of pointer analysis are extant: alias analysis identifies
pairs of address expressions that both hold the address of a given
location. Points-to analysis identifies locations whose addresses are
held by pointers. May- and must- variants of both exist. This paper
restricts itself to points-to analysis.


Points-to relations are computed by identifying locations corresponding
to the left- and right-hand sides of a pointer assignment and taking
their cartesian product~\cite{%
Emami.M.Ghiya.R.Hendren.LJ:1994:Context-sensitive-interprocedural-points-to,%
Kanade.A.Khedker.UP.Sanyal.A:2005:Heterogeneous-Fixed-Points}. 
The points-to pairs of locations that are modified are removed.
May-points-to information at $n$ contains the points-to pairs that hold
along some path reaching $n$ whereas must-points-to information 
contains the pairs that hold along every path reaching $n$
(hence a pointer can have at most one pointee)~\cite{%
Kanade.A.Khedker.UP.Sanyal.A:2005:Heterogeneous-Fixed-Points}. 
Fig.~\ref{fig:intra.pta.conventional} provides an example of flow-sensitive
points-to analysis.  For this example, an inclusion-based flow-insensitive 
analysis~\cite{Andersen.LO:1994:Program-Analysis-and}
concludes that \pt{p}{r}, \pt{p}{s}, \pt{q}{r}, \pt{r}{s}, \pt{s}{r}
hold at all program points.
An equality-based flow-insensitive analysis~\cite{Steensgaard.B:1996:Points-to-Analysis-in}
additionally computes \pt{q}{s}.


\section{Lazy Pointer Analysis}
\label{sec:lazy.pta}

We consider the four basic pointer assignment statements: $x = \&y$,
$x = y$, $x = *y$, $*x = y$ using which other pointer assignments can
be rewritten. We also assume a \text{{\em use x\/}} statement to model
other uses of pointers (such as in conditions).

\subsection{Notation and Basic Definitions}


Let \var denote the set of variables (i.e.\ ``named locations''). Some
of these variables (those in $\pointer \subset \var$) can hold pointers
to members of $\var$. Other members of $\var$ hold non-pointer values.
These include variables of non-pointer type such as {\tt int}. NULL
is similarly best regarded as a member of $\var-\pointer$; finally a
special value `?' in $\var-\pointer$ denotes an undefined location.
This represents the value of an uninitialised pointer declaration,
e.g.\ \mbox{\tt int *x;}. At the moment it is simplest to think of
`?' as being NULL as in Java rather than C, so that indirecting on it
terminates execution (Section~\ref{sec:design.choices} explains this).

Points-to information is a set of pairs \pt{x}{y} where
\text{$x\in\pointer$} is the pointer of the pair and \text{$y\in \var$}
is a pointee of $x$ and is also referred to as the pointee of the pair.
The pair \pt{x}{?} being associated with program point $n$ indicates
that $x$ may not contain a valid address along some potential execution
path from \Start{p} to $n$.

The liveness information for statement $n$ is denoted by the data flow
variables \text{$\lin_n$} and \text{$\lout_n$}, the may-points-to
information is denoted by \text{$\ain_n$} and \text{$\aout_n$}, and
the must-points-to information is denoted by \text{$\uin_n, \uout_n$}.
Instead of being calculated as a mutual fixed point with $\ain_n,
\aout_n$, in our framework \text{$\uin_n, \uout_n$}
are computed afterwards from $\ain_n,
\aout_n$. Note that 
liveness propagates backwards (transfer functions map {\sf\em out\/} to
{\sf\em in\/}) while points-to propagates forwards.

Let \text{$\mathcal{P}(S)$} denote the powerset of $S$. Then \text{$\lvE
= \left\langle\mathcal{P}(\pointer), \supseteq\right\rangle$}
is the lattice of liveness information. Note that this means
that we do not track the liveness of non-pointer variables
because their liveness is not relevant to points-to analysis.
The lattice of may-points-to information is \text{$\aptE =
\left\langle\mathcal{P}(\pointer\times\var), \supseteq\right\rangle$}.
The overall lattice
of our data flow values is the product \text{$\lvE\times\aptE$} having
partial order\footnote{We use the original data flow greatest fixpoint
formulation where $\top$ constitutes the initial value rather
then the abstract-interpretation-style least fixpoint formulation which
iterates from $\bot$.}:
\begin{align}
\begin{split}
\forall 
\langle l_1, a_1  \rangle, 
	\langle l_2, a_2 \rangle \in 
	\lvE\times\aptE, 
	 \langle l_1, a_1 \rangle \sqsubseteq \langle l_2, a_2 \rangle 
	& \Leftrightarrow
	 \left(l_1 \sqsubseteq l_2\right) \wedge \left(a_1 \sqsubseteq a_2\right) 
	\\
	& \Leftrightarrow
	 \left(l_1 \supseteq l_2\right) \wedge \left(a_1 \supseteq a_2\right) 
\end{split}
\end{align}
The $\top$ element of the lattice \text{$\lvE\times\aptE$} is \text{$\langle \emptyset,
\emptyset \rangle$} and the $\bot$ element is \text{$\langle \pointer,
\pointer\times\var\rangle$}.

We use standard algebraic operations on points-to relations:
\begin{itemize}
\item  For a given relation \text{$R \subseteq \pointer\times\var$} and some set $X$, 
       relation {\em application\/} (\text{$R\;X$}) is defined as
	\text{$R \; X   = \{v \mid u\in \!X \wedge \pt{u}{v} \in \!R\} $} 
        and relation {\em restriction\/} (\lrestrict{R}{X}) is defined as
        \text{$ \lrestrict{R}{X}  = \{\pt{u}{v} \in R \mid u\in X\} $}.

\item Given relations \text{$S \subseteq A\times B$} and \text{$T \subseteq
      B\times C$}, {\em relation composition\/} \text{$T\circ S \subseteq
      A \times C$} is defined as 
	\text{$
        T\circ S  = \{ \pt{u}{w} \mid \pt{u}{v} \in S \wedge \pt{v}{w} \in T\}
	$}.
\end{itemize}

However, since \text{$R \subseteq \pointer\times\var$}, we need
to take a little more care formalising $R \circ R$ because of the
mismatch between the sets. We adopt the conventional approach of
using the inclusion map: since \text{$\pointer \subseteq \var$}, by
inclusion of relations we regard the leftmost $R$ as being a subset
of \text{$\var \times \var$} (effectively coercing \text{$\pointer
\times \var$} into \text{$\var \times \var$}). To distinguish it from
the usual composition, we denote it as \text{$R \hat{\circ} R$}.
Note that the result is a subset of \text{$\pointer\times\var$}.

Consider \text{$\var = \{ a, b, c, d, e, f, g, ? \}$} and
\text{$\pointer = \{ a, b, c, d, e\}$}. Let relation \text{$R \subseteq
\pointer\times\var$} be \text{$\{ \pt{a}{b}, \pt{a}{c}, \pt{b}{d},
\pt{c}{e}, \pt{c}{g}, \pt{d}{a},\pt{d}{g}, \pt{e}{?} \}$}. Consider
\text{$Z = \{a, c \}$}. Then:
\begin{eqnarray*}
R \; Z & = & \{ b, c, e, g \}
	\\
\lrestrict{R}{Z}
	& = & 
	\{ 
	\pt{a}{b},
	\pt{a}{c}, 
	\pt{c}{e},
	\pt{c}{g}
	\}
	\\
R^2 & = & R \,\hat{\circ} R \; = \; 
	\{
	\pt{a}{d},
	\pt{a}{e},
	\pt{a}{g},
	\pt{b}{a},
	\pt{b}{g},
	\pt{c}{?}, 
	\pt{d}{b},
	\pt{d}{c}
	\}
\end{eqnarray*}

\subsection{What is Lazy Pointer Analysis?}
\label{sec:lazy.pta.definition}

We formulate liveness and points-to analysis so that points-to
information is computed relative to liveness. In particular a points-to
pair for a pointer is generated only if the pointer is live. Hence
{\tt \{x=\&y; return 3;\}} and {\tt \{x=\&z; return 3;\}} calculate
the same (empty) points-to information as {\tt x} is dead after the
assignment. Further, liveness information is similarly computed relative
to points-to information, using strong
liveness~\cite{Nielson.F.Nielson.HR.Hankin.C:1998:Principles-of-Program}
instead of the more common simple liveness. In strong liveness, the
variables read in an assignment statement are considered live only if
any of the variables defined by the statement are known to be live.
In simple liveness all variables that are read are considered live
regardless of the liveness of the variables that are defined. This
formulation directly allows a joint liveness-and-points-to analysis
Further,
\begin{itemize}
\item The propagation of points-to information is {\em sparse\/} in the
      CFG; a points-to pair \pt{x}{y} is propagated only along those live
      ranges of $x$ that include the statement in which this pair is
      generated. In contrast, the propagation of liveness information
      is {\em dense\/} because it is propagated to all possible program
      points.

\item Must points-to information is computed incrementally from the may
      points-to information without requiring an interdependent fixed
      point computation. This is quite unlike~\cite{%
Emami.M.Ghiya.R.Hendren.LJ:1994:Context-sensitive-interprocedural-points-to,%
Kanade.A.Khedker.UP.Sanyal.A:2005:Heterogeneous-Fixed-Points}. 
\end{itemize}

We use the example of Fig.~\ref{fig:intra.pta.conventional} as a
motivating example for our intraprocedural formulation and make the
following observations:
\begin{itemize}
\item $p$ is live at the exit of 2 because of its use in 3 and 4. $q$
      becomes live because it is used in defining $p$ in 2 and $p$ is
      live at the exit of 2. Hence \pt{q}{r} should be generated in 1
      and should be propagated everywhere except in 7 where $q$ is not
      live.
\item Since \pt{q}{r} holds in 2, \pt{p}{r} should be generated and
      should be propagated to only 3 and 4 because $p$ is not live
      anywhere else.
\item $r$ becomes live in 3 because \pt{p}{r} holds in 3. Its liveness
      is propagated to 1, 2, and 6. It is not live in 4, 5, and 7.
      \pt{r}{s} should be generated in 6 and should be propagated to 2
      and 3 but not beyond because $r$ is not live beyond 2 and 3.
\item $s$ is not live anywhere hence \pt{s}{r} should be not generated.
\end{itemize}

\subsection{Defining Lazy Pointer Analysis}
\label{sec:lazy.pta.formulation}


Fig.~\ref{fig:lazy.pta.dfe} provides the data flow equations for lazy
pointer analysis. They resemble the standard data flow equations of
liveness analysis and pointer
analyses~\cite{Khedker.U.Sanyal.A.Karkare.B:2009:Data-Flow-Analysis}.
However, there are three major differences: 
\begin{itemize}
\item liveness and may-points-to analyses depend on each other (bi-directional),
\item lazy computation and sparse propagation are directly captured in the
      equations, and
\item must-points-to information is computed from may-points-to
      information (i.e.\ fixed-point computation is not performed for must
      points-to analysis).
\end{itemize}
The initial value ($\top$ of the corresponding lattices)
used for computing the fixed point is $\emptyset$
for both liveness and may-points-to analyses.
For liveness \boundary is $\emptyset$
and defines $\lout_{\End{p}}$; whereas
for points-to analysis,
\boundary is \text{$\lin_n\times\{?\}$} and defines $\ain_{\Start{p}}$.
This reflects that no pointer is live on exit or holds a valid address
on entry to a procedure.

\subsubsection{Extractor Functions.}
The flow functions occurring in Equations~(\ref{eq:lin.exhaustive})
and~(\ref{eq:aout.exhaustive}) use {\em extractor functions\/} $\Def_n$,
$\Kill_n$, $\Ref_n$, and $\Pointee_n$ which extract the relevant pointer
variables for statement $n$ from the incoming pointer information
$\ain_n$. These extractor functions are inspired by similar functions 
in~\cite{%
Emami.M.Ghiya.R.Hendren.LJ:1994:Context-sensitive-interprocedural-points-to,%
Kanade.A.Khedker.UP.Sanyal.A:2005:Heterogeneous-Fixed-Points}. 

$\Def_n$ examines the left hand side an assignment statement to find
the pointer variables which may be defined by the statement to
hold new addresses. $\Pointee_n$ computes potential pointees by
examining the right hand side of $n$. Thus the new points-to pairs
generated for statement $n$ are \text{$\Def_n\times\Pointee_n$}
(Equation~\ref{eq:aout.exhaustive}). Sparse propagation of points-to
pairs is ensured by restricting the collected points-to pairs to
live pointers. $\Ref_n$ computes the variables that become live in
statement $n$. Condition \text{$\Def_n\cap\lout_n$} ensures that
$\Ref_n$ computes strong liveness rather than simple liveness. As an
exception to the general rule, $x$ is considered live in statement
\text{$*x=y$} regardless of whether the pointees of $x$ are live
otherwise, the pointees of $x$ would not be discovered. For example,
given \text{\tt\{x=\&a; y=3; *x=y; return;\}}, \pt{x}{a} cannot be
discovered unless $x$ is marked live. Hence liveness of $x$ cannot
depend on whether the pointees of $x$ are live. By contrast, statement
\text{$y=*x$} uses the liveness of $y$ to determine the liveness of $x$.

\begin{figure}[!t]
Given relation $R\subseteq \pointer\times\var$ (either $\ain_n$ or $\aout_n$)
we first define an auxiliary extractor function
\begin{align}
  \must(R) & 
	= \bigcup_{x\in \pointer} \{x\} \times
  \left\{\begin{array}{c@{\hspace{1em}}l}
     \var 
	& \left(\lrestrict{R}{\{x\}} = \emptyset \right)\vee
	  \left(\lrestrict{R}{\{x\}} = \{ \pt{x}{?}\}\right)
		\\
     \{y\} 
	& 
	  \left(\lrestrict{R}{\{x\}} = \{ \pt{x}{y}\}\right) \wedge
	  (y \neq ?)
	\\
     \emptyset & \mbox{otherwise}
  \end{array}\right.
	\label{eq:must.new.def}
\end{align}
\[
\begin{array}{|c|c|c|c|c|c|}
\hline
\multicolumn{6}{|c|}{%
\text{Extractor functions for statement $n$}
}
\\ 
\multicolumn{6}{|c|}{%
\text{Notation: we assume that $x, y \in \pointer$ and $a\in \var$.
$A$ abbreviates $\ain_n$.}\rule[-.5em]{0em}{1.5em}
}
\\ \hline
\multirow{2}{*}{Stmt.} 
	& \multirow{2}{*}{$\Def_n$}
	& \multirow{2}{*}{$\Kill_n$}
	& \multicolumn{2}{c|}{\Ref_n}
	& \multirow{2}{*}{$\Pointee_n$}
	\\ \cline{4-5}
	&
	&
	& \text{if } \Def_n \cap \lout_n \neq \emptyset\rule{0em}{1em}
	& \text{Otherwise}
	&
		\\ \hline\hline
\text{\em use\/} \; x
	& \emptyset
	& \emptyset
	& \rule[-.5em]{0em}{1.5em}
	 \{ x\}
	& \rule[-.5em]{0em}{1.5em}
	 \{ x\}
	& \emptyset
		\\ \hline
x = \& a
	& \{ x\}
	& \{ x\}
	& \rule[-.5em]{0em}{1.5em}
	 \emptyset
	& \rule[-.5em]{0em}{1.5em}
	 \emptyset
	& \{ a\}
		\\ \hline
x = y
	& \{ x\}
	& \{ x\}
	& \{y\}  
	& \emptyset
	& \rule[-.5em]{0em}{1.5em}
	 A\{y\} 
		\\ \hline
x = *y 
	& \{ x\}
	& \{ x\}
	& \{ y \} \cup (A\{y\}) \cap \pointer \;\;
	& \emptyset
	& \rule[-.6em]{0em}{1.75em}
		A^2 \{y\}  
		\\ \hline
* x = y 
	& (A\{x\}) \cap \pointer \;
	 \rule[-.5em]{0em}{1.5em}
	& (\must(A)\{x\}) \cap \pointer 
	&  \{ x, y\}
	&  \{ x\}
	& A\{y\}
		\\ \hline
\text{other} \rule[-.25em]{0em}{1.25em}
	& \emptyset
	& \emptyset
	& \emptyset
	& \emptyset
        &
		\\ \hline
\end{array}
\]

\begin{eqnarray}
\begin{array}{c}
	\text{Domains:}\rule[-3.3em]{0em}{4em}
\end{array}
	&
	&
\begin{array}{|rl|rl|}
\hline
\multicolumn{2}{|c|}{\text{Extractor Functions}}
	& \multicolumn{2}{c|}{\text{Data Flow Values}}
	\\ \hline\hline
\Pointee_n  \subseteq  & \var\; \;& \lin_n, \lout_n \subseteq &  \pointer\; \;
		\\ \hline
\Kill_n, \Def_n, \Ref_n  \subseteq  & \pointer\; \;&
\ain_n, \aout_n  \subseteq & \pointer \times \var\;\; 
		\\ \hline
\end{array}
\nonumber \\
\nonumber \\
\lout_n & = & 
	\left\{
		\begin{array}{c@{\ \ \ \ }l}
			\emptyset 	& n \text{ is } \End{p}
				\\
			\displaystyle\bigcup_{s \in \Succ(n)} \lin_s
				& \text{otherwise}
		\end{array}
	\right.
	\label{eq:lout.exhaustive}
	\\
\lin_n & = & 
                \left(\lout_n - \Kill_n\right) \cup \Ref_n 
	\label{eq:lin.exhaustive}
	\\
\ain_n & = & 
	\left\{ \renewcommand{\arraystretch}{1.2}
		\begin{array}{c@{\ \ \ \ }l}
			\lin_n \!\times\! \{?\}
                                & n \text{ is } \Start{p}
				\\
		\left.
		\left(\displaystyle\bigcup_{p \in \pred(n)} \aout_p\right)
		\right|_{\mbox{$\lin_n$}}
				& \text{otherwise}
		\end{array}
	\right.
	\label{eq:ain.exhaustive}
	\\
\aout_n & = & 
\lrestrict{
                \left(\left(\ain_n  -  \left(\Kill_n \!\times\! \var\right)\right) \cup 
			\left( \Def_n \!\times\! \Pointee_n\right)\right)
}{\mbox{$\lout_n$}}
	\label{eq:aout.exhaustive}
\end{eqnarray}
\caption{Intraprocedural formulation of lazy pointer analysis.}
\label{fig:lazy.pta.dfe}
\end{figure}

$\Kill_n$ identifies pointer variables that are definitely modified by
the execution of statement $n$. This information is used for killing
liveness as well as points-to information. For statement \text{$*x=y$},
$\Kill_n$ depends on $\ain_n$ which is filtered using the function
\must. The filtering criteria ensures that when no points-to information
for $x$ is available, we conservatively assume that all pointers are
modified by statement \text{$*x=y$}. This is consistent with the
initial values of may-points-to information and liveness both of which
are $\emptyset$. Given some points-to information for $x$, \must
uses the number of pointees to determine whether to perform a {\em weak
update\/} or a {\em strong update\/}: when $x$ has multiple pointees
we cannot be certain which one will be modified because $x$ points to
different locations along different execution paths reaching $n$. In
this case we employ weak update which does not allow any data flow
information to be killed. By contrast, when $x$ has a single pointee
{\em other than `?'}, it indicates that $x$ points to the same location
along all execution paths reaching $n$ and a strong update can be
performed.\footnote{ Note that this conclusion is only possible because
\boundary is \text{$\lin_n\times\{?\}$}. This value of \boundary ensures
that if there is a definition-free path from the \Start{p} statement
to statement $n$, we will get \pt{x}{?} at $n$ and so a solitary
pair \pt{x}{z} also reaching $n$ is not incorrectly treated as a
must-points-to pair.}

\subsubsection{Motivating Example Revisited.}
\begin{figure}[t]
\renewcommand{\pt}[2]{\text{$(#1,#2)$}}
\renewcommand{\dfv}[2]{\text{$\{\protect#1\}_L,\{\protect#2\}_A$}}
\begin{tabular}{c|c}
\begin{tabular}{c}
\begin{pspicture}(0,0)(59,82)
\putnode{n1}{origin}{32}{79}{\psframebox{$q = \&r$}}
\putnode{w}{n1}{8}{0}{1}
\putnode{n2}{n1}{0}{-15}{\psframebox{$p=q$}}
\putnode{w}{n2}{7}{0}{2}
\putnode{n3}{n2}{-12}{-16}{\psframebox{$p=*p$}}
\putnode{w}{n3}{7}{0}{\;3}
\putnode{n4}{n3}{0}{-16}{\psframebox{print $p$}}
\putnode{w}{n4}{7}{0}{\,4}
\putnode{n5}{n2}{12}{-24}{\psframebox{$s=q$\rule{0em}{.5em}}}
\putnode{w}{n5}{7}{0}{5}
\putnode{n6}{n4}{12}{-13}{\psframebox{$r=\&s$\rule{0em}{.3em}}}
\putnode{w}{n6}{8}{0}{6}
\putnode{n7}{n6}{0}{-15}{\psframebox{\white$p=*p$}\;}
\putnode{w}{n7}{8}{0}{7}
\ncline{->}{n1}{n2}
\ncline{->}{n2}{n3}
\ncline{->}{n2}{n5}
\ncline{->}{n3}{n4}
\ncline{->}{n4}{n6}
\ncline{->}{n5}{n6}
\ncline{->}{n6}{n7}
\ncloop[angleA=270,angleB=90,offset=2,loopsize=-21,arm=2.5,linearc=.5]{->}{n6}{n2}
\psset{framesep=0,linestyle=none,fillstyle=solid,fillcolor=lightgray}
\psset{framearc=0} 
\putnode{w}{n7}{0}{5}{\psframebox{\dfv{}{}}}
\putnode{w}{n6}{0}{-5}{\psframebox{\dfv{q}{\pt{q}{r}}}}
\putnode{w}{n6}{2}{5}{\psframebox{\dfv{q}{\pt{q}{r}}}}
\putnode{w}{n5}{0}{5}{\psframebox{\dfv{q}{\pt{q}{r}}}}
\putnode{w}{n5}{0}{-5}{\psframebox{\dfv{q}{\pt{q}{r}}}}
\putnode{w}{n4}{0}{5}{\psframebox{\dfv{p,q}{\pt{q}{r}}}}
\putnode[r]{w}{n4}{2}{-5}{\psframebox{\dfv{q}{\pt{q}{r}}}}
\putnode{w}{n3}{0}{5}{\psframebox{\dfv{p,q}{\pt{p}{r},\pt{q}{r}}}}
\putnode{w}{n3}{0}{-5}{\psframebox{\dfv{p,q}{\pt{q}{r}}}}
\putnode{w}{n2}{0}{5}{\psframebox{\dfv{q}{\pt{q}{r}}}}
\putnode{w}{n2}{0}{-5}{\psframebox{\dfv{p,q}{\pt{p}{r},\pt{q}{r}}}}
\putnode{w}{n1}{0}{-5}{\psframebox{\dfv{q}{\pt{q}{r}}}}
\end{pspicture}
\end{tabular}
&
\begin{tabular}{c}
\begin{pspicture}(0,0)(59,82)
\putnode{n1}{origin}{33}{79}{\psframebox{$q = \&r$}}
\putnode{w}{n1}{8}{0}{1}
\putnode{n2}{n1}{0}{-15}{\psframebox{$p=q$}}
\putnode{w}{n2}{7}{0}{2}
\putnode{n3}{n2}{-12}{-16}{\psframebox{$p=*p$}}
\putnode{w}{n3}{7}{0}{\;3}
\putnode{n4}{n3}{0}{-16}{\psframebox{print $p$}}
\putnode{w}{n4}{7}{0}{\,4}
\putnode{n5}{n2}{12}{-24}{\psframebox{$s=q$\rule{0em}{.5em}}}
\putnode{w}{n5}{7}{0}{5}
\putnode{n6}{n4}{12}{-13}{\psframebox{$r=\&s$\rule{0em}{.3em}}}
\putnode{w}{n6}{8}{0}{6}
\putnode{n7}{n6}{0}{-15}{\psframebox{\white$p=*p$}\;}
\putnode{w}{n7}{8}{0}{7}
\ncline{->}{n1}{n2}
\ncline{->}{n2}{n3}
\ncline{->}{n2}{n5}
\ncline{->}{n3}{n4}
\ncline{->}{n4}{n6}
\ncline{->}{n5}{n6}
\ncline{->}{n6}{n7}
\ncloop[angleA=270,angleB=90,offset=2,loopsize=-21,arm=2.5,linearc=.5]{->}{n6}{n2}
\psset{framesep=0,linestyle=none,fillstyle=solid,fillcolor=lightgray}
\psset{framearc=0} 
\putnode{w}{n7}{0}{5}{\psframebox{\dfv{}{}}}
\putnode{w}{n6}{0}{-5}{\psframebox{\dfv{q,r}{\pt{q}{r},\pt{r}{s}}}}
\putnode{w}{n6}{2}{5}{\psframebox{\dfv{q}{\pt{q}{r}}}}
\putnode{w}{n5}{2}{5}{\psframebox{\dfv{q}{\pt{q}{r}}}}
\putnode{w}{n5}{2}{-5}{\psframebox{\dfv{q}{\pt{q}{r}}}}
\putnode{w}{n4}{-3}{5}{\psframebox{\dfv{p,q}{\pt{p}{s},\pt{q}{r}}}}
\putnode[r]{w}{n4}{2}{-5}{\psframebox{\dfv{q}{\pt{q}{r}}}}
\putnode{w}{n3}{3}{5}{\psframebox{\dfv{p,q,r}{\pt{p}{r},\pt{q}{r},\pt{r}{s}}}}
\putnode{w}{n3}{-3}{-5}{\psframebox{\dfv{p,q}{\pt{p}{s},\pt{q}{r}}}}
\putnode{w}{n2}{0}{5}{\psframebox{\dfv{q,r}{\pt{q}{r},\pt{r}{s}}}}
\putnode{w}{n2}{0}{-5}{\psframebox{\dfv{p,q,r}{\pt{p}{r},\pt{q}{r},\pt{r}{s}}}}
\putnode{w}{n1}{0}{-5}{\psframebox{\dfv{q,r}{\pt{q}{r}}}}
\end{pspicture}
\end{tabular}
\\
 First round of liveness and points-to 
   &  Second round of liveness and points-to 
\end{tabular}
\caption{Intraprocedural lazy points-to analysis of the program in 
         Fig.~\ref{fig:intra.pta.conventional}. Shaded boxes show the 
	liveness and points-to information suffixed by $L$ and $A$ 
	respectively.}
\label{fig:intra.pta.result}
\end{figure}
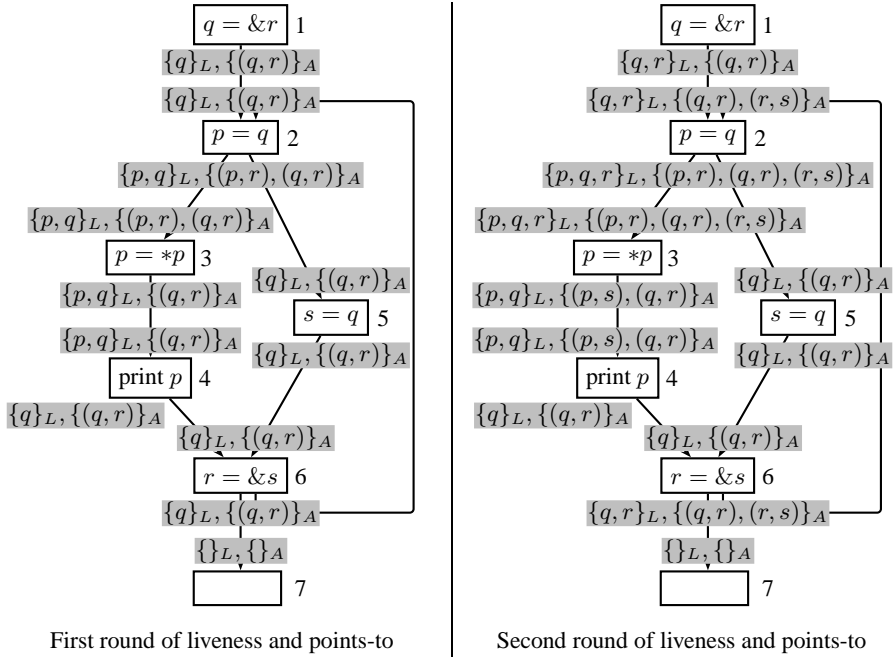

Fig.~\ref{fig:intra.pta.result} gives the result of
lazy pointer analysis for our motivating example of
Fig.~\ref{fig:intra.pta.conventional}. After the first round of
liveness analysis followed by points-to analysis, we discover pair
\pt{p}{r} in $\ain_3$. Thus $r$ becomes live requiring a second round
of liveness analysis. This then enables discovering the points-to pair
\pt{r}{s} in node 6. 
Note that the result is consistent with the observation
made in Section~\ref{sec:lazy.pta.definition} towards the end.
A comparison with the result of the default may-point-to analysis
(Fig.~\ref{fig:intra.pta.conventional}) shows that
our analysis eliminates many redundant points-to pairs.

\subsubsection{Independent Must-points-to Analysis is Redundant.}

The explanation of $\Kill_n$ and 
\must highlights why must-points-to analysis need not be performed
explicitly. With the provision of \text{$\lin_n\times\{?\}$} as
\boundary, if we have a single points-to pair \pt{x}{y} with \text{$y\neq?$}
for pointer $x$ in $\ain_n$ or $\aout_n$, it is guaranteed that $x$
must point to $y$. Conversely multiple points-to pairs associated with
a given variable means that the must-points-to information for this
variable is empty. Hence must-points-to information can be extracted
from may-points-to information by \text{$\uin_n = \must(\ain_n)$} and
\text{$\uout_n = \must(\aout_n)$}. Note that generally $\uin_n \subseteq
\ain_n$ and $\uout_n \subseteq \aout_n$; the only exception would be
for nodes that are not reached by the analysis because no pointer has
been found to be live. For such nodes $\uout_n, \uout_n$ are $\pointer
\times \var$ whereas $\aout_n, \aout_n$ are $\emptyset$; this matches
previous frameworks and is necessary to make \must anti-monotonic
(see~(\ref{eq:monotonicity.5}) in Section~\ref{sec:monotonicity}) as is
required by the the data flow equations in Fig.~\ref{fig:lazy.pta.dfe}.
Note that these definitions avoid the interdependent fixed-point
computation
of~\cite{Kanade.A.Khedker.UP.Sanyal.A:2005:Heterogeneous-Fixed-Points,%
Khedker.U.Sanyal.A.Karkare.B:2009:Data-Flow-Analysis}.

\subsection{Design Choices in Formulating Lazy Pointer Analysis}
\label{sec:design.choices}

\begin{figure}[t]
\hspace*{-2mm}
\begin{tabular}{@{}c|c}
\setlength{\codeLineLength}{55mm}%
\begin{tabular}{@{}c}
	\codeLine{1}{int f(int a, int b)}{white}
	\codeLine{1}{\OB \,int $*x$, $*y = \&a$; \hspace*{10mm}// x = `?'}{white}
	\codeLine{2}{int $c = a*b$;}{white}
	\codeLine{1}{p: $\;*x = 5$; \hspace*{15mm}// Illegal write}{white}
	\codeLine{1}{q: $\;*y = 6$; \hspace*{2mm}// Should kill avail($a*b$)}{white}
	\codeLine{2}{return $c+(a*b)$;}{white}
	\codeLine{1}{\CB}{white}
	\end{tabular}
& 
\begin{minipage}{65mm}
\raggedright
\begin{itemize}
\item \text{$\Def_{p} = \emptyset$}. 
\item Liveness of $y$ is killed.
\item Points-to information of $y$ is empty
      (monotonicity of $\must$ requires this).
\item Statement $q$ cannot kill availability of $a*b$.
\item The return value will be mis-optimised into \text{$c+c$}.
\end{itemize}
\end{minipage}
\end{tabular}
\caption{Motivating the need for a sanity check for programs with undefined behaviour.}
\label{fig:sanity.check}
\end{figure}

We have chosen not to compute liveness of non-pointer variables (keeping
them in $\var-\pointer$) primarily for simplicity and efficiency of
implementation. While we could in principle regard these as members of
$\pointer$ which may be live but can never point to anything, this does not
help discovering points-to information.

In our formulation, data flow value `?' plays an ambiguous role:
it represents an uninitialised pointer value, but it is left unclear
whether this means ``points to some variable in $\var$'' or ``may be
a wild pointer as in C''. We have formulated the analysis as if for
Java; an assignment \mbox{\tt *x=y} can be assumed only to write to a
non-`?' member of the points-to set of {\tt x}---writes to `?' raise an
exception.
In C however, writes via uninitialised pointers can write to any
location in memory, including all user variables. There are two ways to
address this. Firstly, we can treat `?' as not standing for a single
pointer value, but instead being a set of locations including $\var$.
While formally most correct, this requires modification of data flow
equations in Fig.~\ref{fig:lazy.pta.dfe}, for example to $\must$.

Alternatively, and this is the course we have followed (because it
gives more optimisation opportunities), we can optimise a C-like
program as if no dereferences or assignments via invalid pointer
may occur at run-time, but add a ``sanity check'' to stop invalid
optimisations when an illegal pointer assignment necessarily happens.
This is possible because the semantics of both C and Java is that code
after an assignment via an illegal pointer is effectively unreachable.
In Java an exception is raised so the following code is not reached.
In C the behaviour is ``undefined'' so the code can do what it wants,
which includes ``being mis-optimised'' as a special case. However, the
possibility of mis-optimisations arising out of a wild write can be
detected.
We observe that such a situation cannot arise in programs in which every
pointer is defined along {\em some} path before being dereferenced---it
is just the guaranteed dereference of `?' which causes the problem.
Corollary~\ref{cor.def.not.empty} in Section~\ref{sec:sufficiency}
asserts this formally. Hence as the final step in our analysis we
perform a sanity check: there must be no statement \text{$*x=y$}
for which \text{$\Def_n$} is $\emptyset$---otherwise optimisation
is disabled. Fig.~\ref{fig:sanity.check} provides an example that
motivates this sanity check as a form of data-flow-anomaly warning (for
indirect assignments via a variable with no valid pointees).


In our formulation, liveness is generated from points-to information as
indicated by the presence of $\ain_n\{y\}$ in $\Ref_n$ for the statement
$x=*y$ in Fig.~\ref{fig:lazy.pta.dfe}. This leads to imprecision in
liveness information which in turn leads to imprecision in points-to
information as illustrated in Fig.~\ref{fig:indirect.exmp}.
This imprecision can be avoided by exploiting
the mutual dependence of liveness and points-to information: i.e.\
propagation of indirect liveness should also be restricted to
appropriate points-to propagation paths---in the same way that
propagation of points-to pairs is restricted to liveness paths.
We omit this formulation (which is not used in the implementation)
for space reasons.

\begin{figure}[t]
\begin{center}
\begin{tabular}{@{}cc}
\begin{tabular}{@{}c@{}}
\begin{pspicture}(0,0)(38,32)
\putnode{n1}{origin}{20}{30}{\psframebox{\white$a = \& b$}}
\putnode{w}{n1}{-8}{0}{1}
\putnode{n2}{n1}{-9}{-12}{\psframebox{$a = \& b$}}
\putnode{w}{n2}{-8}{0}{2}
\putnode{n3}{n1}{9}{-8}{\psframebox{$b = \& e$}}
\putnode{w}{n3}{-8}{0}{3}
\putnode{n4}{n3}{0}{-8}{\psframebox{$a = \& c$}}
\putnode{w}{n4}{-8}{0}{4}
\putnode{n5}{n2}{9}{-12}{\psframebox{$d = * a$}}
\putnode{w}{n5}{-8}{0}{5}
\putnode{n6}{n5}{0}{-8}{\psframebox{$use \;\;d $}}
\putnode{w}{n6}{-8}{0}{6}
\ncline{->}{n1}{n3}
\ncline{->}{n1}{n2}
\ncline{->}{n3}{n4}
\ncline{->}{n2}{n5}
\ncline{->}{n4}{n5}
\ncline{->}{n5}{n6}
\end{pspicture}
\end{tabular}
&
\begin{minipage}{78mm}
\begin{itemize}
\item Since $a$ is live at the exit of 2 and 3, pairs \pt{a}{b} and
      \pt{a}{c} are generated which causes $b$ and $c$ to be marked live
      in 5. Hence \text{$\lout_2 = \lout_4 = \{a, b, c\}$}.

\item However, $b$ is not live along path \text{1-3-4-5} because        
      \text{$\pt{a}{b} \not \in \aout_4$}. Similarly, $c$ is not live   
      along path \text{1-2-5} because \text{$\pt{a}{c} \not \in \aout_2$}. 

\item Due to this imprecision in liveness, we generate the pair
      \pt{b}{e} in 3 which is then propagated to 5. This is spurious
      because there is no use of $b$ anywhere along this path.
\end{itemize}
\end{minipage}
\end{tabular}
\end{center}
\caption{Imprecision in lazy pointer analysis due to indirect liveness.}
\label{fig:indirect.exmp}
\end{figure}

\section{Properties of Lazy Pointer Analysis}
\label{sec:properties}

In this section we show that lazy pointer analysis is monotonic,
sparse, and discovers all pointees of a pointer variable where 
it is used.
Proofs are provided in Appendix~\ref{app:proofs}. 

\subsubsection{Monotonicity of Lazy Pointer Analysis}
\label{sec:monotonicity}
The extractor functions $\Def_n$, $\Pointee_n$, $\Kill_n$ (and \must)
use points-to
information. Besides, $\Ref_n$ uses liveness information also. In order
to argue about monotonicity, we parameterise the extractor functions
with the required information and drop the subscript $n$.

Recall that the lattice of points-to information is \text{$\aptE =
\left\langle\mathcal{P}(\pointer\times\var), \supseteq\right\rangle$}.
The following
results hold \text{$\forall v \in \var$} and \text{$\forall a_1, a_2 \in
\aptE$} such that \text{$a_1 \sqsubseteq a_2$} (i.e.\ $a_1 \supseteq a_2$):
\begin{align}
  a_1\{v\} & \supseteq a_2\{v\}
	& \text{(follows from the definition)}
	\label{eq:monotonicity.1}
	\\
  (a_1\circ a_1)\{v\} & \supseteq (a_2\circ a_2)\{v\}
	& \text{(follows from the definition)}
	\label{eq:monotonicity.2}
	\\
  \Def\,(a_1) & \supseteq \Def\,(a_2)
	& \text{(follows from~(\ref{eq:monotonicity.1}))}
	\label{eq:monotonicity.3}
	\\
  \Pointee\,(a_1) & \supseteq \Pointee\,(a_2)
	& \text{(follows from~(\ref{eq:monotonicity.2}))}
	\label{eq:monotonicity.4}
	\\
  \left(\must(a_1)\right)\{v\} & \subseteq 
				\left(\must(a_2)\right)\{v\} 
	& \text{(follows from~(\ref{eq:must.new.def}) and~(\ref{eq:monotonicity.1}))}
	\label{eq:monotonicity.5}
\end{align}


\begin{theorem}
Define the liveness flow function \text{$f_L : \lvE\times\aptE
\mapsto \lvE$} (Equation~\ref{eq:lin.exhaustive}) and the may-points-to
flow function \text{$f_A : \lvE\times\aptE \mapsto \aptE$}
(Equation~\ref{eq:aout.exhaustive}) as:
\begin{align*}
f_L (l,a) & = (l - \Kill\,(a)) \cup \Ref\,(l,a)
	\\
f_A (l,a) & = \bigrestrict{\left(a - (\Kill\,(a)\times \var)\right) 
	\cup \left(\Def\,(a)\times\Pointee\,(a)\right)}{l}
\end{align*}
Then, $f_L$ and $f_A$ are monotonic.
\end{theorem}

\subsubsection{Sparseness of Lazy Pointer Analysis}
\label{sec:sparseness}

Lazy pointer analysis is a form of sparse data flow analysis: the
analysis is only done on live ranges rather than everywhere (contrast
the previous use of the term to mean ``along def-use chains'').

Equations~(\ref{eq:lout.exhaustive}) and~(\ref{eq:lin.exhaustive})
identify {\em liveness paths\/} for variables representing control
flow paths in the program along which variables are live. For a given
variable $x$, a liveness path is defined as a maximal sequence of
statements \text{$s_1, s_2, \cdots, s_k$} satisfying the following
conditions:
\begin{itemize}
\item  \text{$x \in \Ref_{s_k}$}.
\hfill	(L1)
\item  \text{$s_{i+1} \in \Succ(s_i), 1 \leq i < k$}.
\hfill	(L2) 
\item  $x \not\in \Kill_{s_i}, 
       x \in \lin_{s_i}, 
	1 < i \leq k$. 
\hfill	(L3) 
\item  $x \in \lout_{s_i}, 
	1 \leq i < k$.
\hfill	(L4) 
\end{itemize}
(L1) represents generation of liveness of $x$, (L2) insists on
the sequence being a control flow path, while (L3) and (L4)
ensure that the sequence is a modification free path. 

Equations~(\ref{eq:ain.exhaustive}) and~(\ref{eq:aout.exhaustive})
identify {\em propagation paths\/} for points-to pairs representing
control flow paths in the program along which points-to pairs are
propagated. For a given points-to pair \pt{x}{y}, a propagation path is
defined as a maximal sequence of statements \text{$s_1, s_2, \cdots,
s_k$} satisfying the following conditions:
\begin{itemize}
\item  \text{$\pt{x}{y} \in \left(\Def_{s_1} \times \Pointee_{s_1}\right)$}.
	\hfill (A1) 
\item  \text{$s_{i+1} \in \Succ(s_i), 1 \leq i < k$}.
	\hfill (A2) 
\item  $\pt{x}{y} \in \ain_{s_i},  
       x \not\in \Kill_{s_i}, 
       x \in \lin_{s_i}, 
	1 < i \leq k$. 
	\hfill (A3) 
\item  $\pt{x}{y} \in \aout_{s_i}, 
       x \in \lout_{s_i}, 
	1 \leq i < k$.
	\hfill (A4) 
\end{itemize}
(A1) represents generation of the pair \pt{x}{y}, (A2) ensures that the
sequence is a control flow path, while (A3) and (A4) ensure propagation
along a modification free path. 

\begin{theorem}
Every propagation path for a points-to pair \pt{x}{y} is a suffix of some 
liveness path for $x$.
\end{theorem}

\subsubsection{Sufficiency of Lazy Pointer Analysis}
\label{sec:sufficiency}
At the program point of every use of a
pointer variable, lazy pointer analysis discovers all pointees of
the pointer variable. We first observe a useful relationship between
$\Kill_n$ and $\Def_n$.

\begin{lemma}
\label{lemma.kill.def}
$\left((\Kill_n \neq \emptyset ) \wedge (\Kill_n \neq \pointer)\right) 
       \Rightarrow \left(\Kill_n = \Def_n\right).
$
\end{lemma}

\begin{theorem}
\label{theorem.live.pt}
If \text{$x\in \pointer$} holds the address of \text{$z \in (\var - \{?\})$}
along some execution path reaching node $n$, then \text{$x \in \Ref_n \Rightarrow 
\pt{x}{z} \in \ain_n$}.
\end{theorem}

\begin{corollary}
\label{cor.def.not.empty}
If all pointer variables are initialised with values of proper types
before they are used, then for every indirect assignment \text{$*x=y$},
\text{$\Def_n \neq \emptyset$}.
\end{corollary}

\section{Interprocedural Lazy Pointer Analysis}
\label{sec:interprocedural}

We use the call-strings method (Section~\ref{sec:ipa.callstrings})
to ensure flow- and context-sensitivity.
Since it is a generic method orthogonal to any particular analysis,
lifting an intraprocedural formulation of an analysis to interprocedural
level is straightforward. In our case, $\lin_n, \lout_n$ and $\ain_n,
\aout_n$ become sets of pairs \text{$\sdfv{\sigma}{a}, a \in \aptE$} and
\text{$\sdfv{\sigma}{a}, l \in \lvE$} at the interprocedural level
where $\sigma$ is a call string reaching node $n$. The final values of
$\ain_n, \aout_n$ are computed by merging the values along all call
strings.

\subsubsection{Terminating Call String Construction.}

We use data flow values~\cite{Khedker.UP.Karkare.B:2008:Efficiency-Precision-Simplicity}
to terminate call-string construction instead of using a precomputed length 
as proposed originally~\cite{Sharir.M.Pnueli.A:1981:Two-Approaches-to}.
This approach discards redundant call strings at \Start{p} and regenerates them at \End{p} for
forward flows  as follows (and the other way round for backward flows):
\begin{itemize}
\item {\em Representation}. If two call strings $\sigma$ and $\sigma'$
	have identical data flow values at \Start{p}, both need not be 
        propagated within the body of $p$ because the data flow values of 
        both $\sigma$ and $\sigma'$ will undergo the same change and 
        will remain identical at \End{p}. 
        More formally, \Out{\Start{p}} is now computed as follows:
\begin{align*}
\Out{\Start{p}} 
	& = \left\{ \rep\left(\sdfv{\sigma}{x}, {p}\right)
		\mid
		\sdfv{\sigma}{x} \in \In{\Start{p}}
		\right\}
	& \text{where,}
	\\
\rep\left(\sdfv{\sigma}{x}, {p}\right) 
	& = 
	\left\{
	\begin{array}{l@{\ \ \ }l}
	\sdfv{\sigma'}{x} & \sdfv{\sigma'}{x} \in \In{\Start{p}},
				|\sigma'| \leq |\sigma|
		\\
	 \sdfv{\sigma}{x} & \text{otherwise}
	\end{array}
	\right.
\end{align*}

\item {\em Regeneration}. At \End{p}, we 
        examine the representation performed at \Start{p}. If $\sigma$
        represents $\sigma'$, the data flow value associated with $\sigma$
        it is copied to $\sigma'$. Thus,
\begin{align*}
\Out{\End{p}} 
	& = 
	\left\{\sdfv{\sigma'}{y} \mid 
		\sdfv{\sigma}{y} \in \In{\End{p}},
		\sdfv{\sigma'}{y} \in 
                  \reg\left(\sdfv{\sigma}{y}, {p}\right)
	\right\}
	\\
\reg\left(\sdfv{\sigma}{y}, {p}\right) 
	& = 
	\left\{ \sdfv{\sigma'}{y} \mid
		\rep\left(\sdfv{\sigma'}{x}, {p}\right) 
		= \sdfv{\sigma}{x} 
	\right\}
\end{align*}
\end{itemize}
Representation partitions call strings into equivalence classes based
on the data flow values associated with them. Regeneration recreates
the represented call string and recovers their values based on the
partitions they belong to.

\newcommand{\ccs}[1]{\text{$\sigma_c\alpha^{#1}$}\xspace}
\newcommand{\acs}{\text{$\sigma_a$}\xspace}

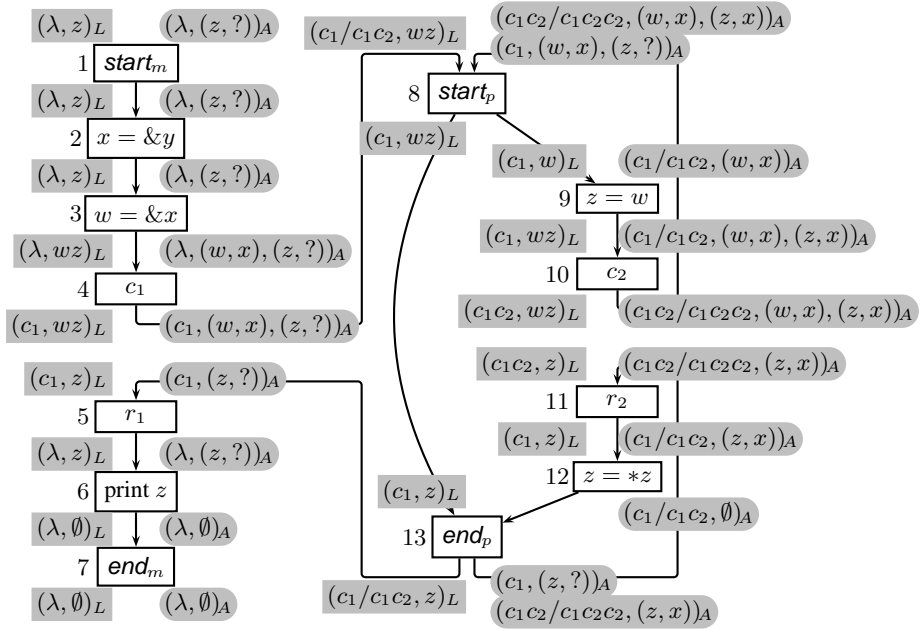
\begin{figure}[!t]
\renewcommand{\pt}[2]{\text{$(#1,#2)$}}
\centering
\begin{tabular}{@{}c}
\begin{pspicture}(0,0)(122,84)
\putnode{entry}{origin}{18}{76}{\psframebox{\Start{m}}}
\putnode{w}{entry}{-7}{0}{$1$}
\putnode{n1}{entry}{0}{-10}{\psframebox{$x = \&y$}}
\putnode{w}{n1}{-8}{0}{$2$}
\putnode{n2}{n1}{0}{-10}{\psframebox{$w = \&x$}}
\putnode{w}{n2}{-8}{0}{$3$}

\putnode{c1}{n2}{0}{-10}{\psframebox{$\;\;\;c_1\;\;\;$}}
\putnode{w}{c1}{-7}{0}{$4$}
\putnode{r1}{c1}{0}{-17}{\psframebox{$\;\;\;r_1\;\;\;$}}
\putnode{w}{r1}{-7}{0}{$5$}
\putnode{n4}{r1}{0}{-10}{\psframebox{print {$z$}}}
\putnode{w}{n4}{-7}{0}{$6$}
\putnode{exit}{n4}{0}{-10}{\psframebox{\End{m}}}
\putnode{w}{exit}{-7}{0}{$7$}
\putnode{sp}{entry}{44}{-4}{\psframebox{\Start{p}}}
\putnode{w}{sp}{-7}{0}{$8$}
\putnode{n5}{sp}{20}{-14}{\psframebox{$z=w$}}
\putnode{w}{n5}{-7}{0}{$9$}
\putnode{c2}{n5}{0}{-10}{\psframebox{$\;\;\;c_2\;\;\;$}}
\putnode{w}{c2}{-8}{0}{$10$}
\putnode{r2}{c2}{0}{-17}{\psframebox{$\;\;\;r_2\;\;\;$}}
\putnode{w}{r2}{-8}{0}{$11$}
\putnode{n6}{r2}{0}{-10}{\psframebox{$z = *z$}}
\putnode{w}{n6}{-8}{0}{$12$}
\putnode{ep}{n6}{-20}{-8}{\psframebox{\End{p}}}
\putnode{w}{ep}{-7}{0}{$13$}
\ncline{->}{entry}{n1}
\ncline{->}{n1}{n2}
\ncline{->}{n2}{n3}
\ncline{->}{n2}{c1}
\ncline{->}{r1}{n4}
\ncline{->}{n4}{exit}
\ncline{->}{sp}{n5}
\ncline{->}{n5}{c2}
\ncline{->}{r2}{n6}
\ncline{->}{n6}{ep}
\nccurve[angleA=240,angleB=120,ncurv=.8]{->}{sp}{ep}
\ncloop[angleA=270,angleB=90,loopsize=-8,linearc=.5,offsetB=1,arm=2.5]{->}{c2}{sp}
\ncloop[angleA=270,angleB=90,loopsize=-27,linearc=.5,offsetA=1,arm=2.5]{->}{ep}{r2}
\ncloop[angleA=270,angleB=90,loopsize=-30,linearc=.5,offsetB=-1,arm=2.5]{->}{c1}{sp}
\ncloop[angleA=270,angleB=90,loopsize=13,linearc=.5,offsetA=-1,arm=2.5]{->}{ep}{r1}
\psset{framesep=.5,linestyle=none,fillstyle=solid,fillcolor=lightgray}
\putnode[r]{w}{entry}{-3}{5}{\psframebox{\dfvl{\lambda}{z}}}
\putnode[r]{w}{n1}{-3}{5}{\psframebox{\dfvl{\lambda}{z}}}
\putnode[r]{w}{n2}{-3}{5}{\psframebox{\dfvl{\lambda}{z}}}
\putnode[r]{w}{c1}{-3}{5}{\psframebox{\dfvl{\lambda}{w z}}}
\putnode[r]{w}{c1}{-3}{-5}{\psframebox{\dfvl{c_1}{w z}}}
\putnode[r]{w}{r1}{-3}{5}{\psframebox{\dfvl{c_1}{z}}}
\putnode[r]{w}{r1}{-3}{-5}{\psframebox{\dfvl{\lambda}{z}}}
\putnode[r]{w}{n4}{-3}{-5}{\psframebox{\dfvl{\lambda}{\emptyset}}}
\putnode[r]{w}{exit}{-3}{-5}{\psframebox{\dfvl{\lambda}{\emptyset}}}
\psset{framearc=2}
\putnode[l]{w}{entry}{3}{5}{\psframebox{\dfva{\lambda}{\pt{z}{?}}}}
\putnode[l]{w}{n1}{3}{5}{\psframebox{\dfva{\lambda}{\pt{z}{?}}}}
\putnode[l]{w}{n2}{3}{5}{\psframebox{\dfva{\lambda}{\pt{z}{?}}}}
\putnode[l]{w}{c1}{3}{5}{\psframebox{\dfva{\lambda}{\pt{w}{x},\pt{z}{?}}}}
\putnode[l]{w}{c1}{3}{-5}{\psframebox{\dfva{c_1}{\pt{w}{x},\pt{z}{?}}}}
\putnode[l]{w}{r1}{3}{5}{\psframebox{\dfva{c_1}{\pt{z}{?}}}}
\putnode[l]{w}{r1}{3}{-5}{\psframebox{\dfva{\lambda}{\pt{z}{?}}}}
\putnode[l]{w}{n4}{3}{-5}{\psframebox{\dfva{\lambda}{\emptyset}}}
\putnode[l]{w}{exit}{3}{-5}{\psframebox{\dfva{\lambda}{\emptyset}}}
\psset{framearc=0}
\putnode[r]{w}{sp}{0}{-6}{\psframebox{\dfvl{c_1}{w z}}}
\putnode[r]{w}{sp}{0}{8}{\psframebox{\dfvl{c_1/c_1c_2}{w z}}}
\putnode[r]{w}{n5}{-4}{5}{\psframebox{\dfvl{c_1}{w}}}
\putnode[r]{w}{n5}{-4}{-5}{\psframebox{\dfvl{c_1}{w z}}}
\putnode[r]{w}{c2}{-4}{-5}{\psframebox{\dfvl{c_1c_2}{w z}}}
\putnode[r]{w}{r2}{-4}{5}{\psframebox{\dfvl{c_1c_2}{z}}}
\putnode[r]{w}{n6}{-4}{5}{\psframebox{\dfvl{c_1}{z}}}
\putnode[r]{w}{ep}{0}{6}{\psframebox{\dfvl{c_1}{z}}}
\putnode[r]{w}{ep}{0}{-8}{\psframebox{\dfvl{c_1/c_1c_2}{z}}}
\psset{framearc=2}
\putnode[l]{w}{sp}{3}{6}{\psframebox{\dfva{c_1}{\pt{w}{x},\pt{z}{?}}}}
\putnode[l]{w}{sp}{3}{10}{\psframebox{\dfva{c_1c_2/c_1c_2c_2}{\pt{w}{x},\pt{z}{x}}}}
\putnode[l]{w}{n5}{0}{5}{\psframebox{\dfva{c_1/c_1c_2}{\pt{w}{x}}}}
\putnode[l]{w}{c2}{0}{5}{\psframebox{\dfva{c_1/c_1c_2}{\pt{w}{x},\pt{z}{x}}}}
\putnode[l]{w}{c2}{0}{-5}{\psframebox{\dfva{c_1c_2/c_1c_2c_2}{\pt{w}{x},\pt{z}{x}}}}
\putnode[l]{w}{r2}{0}{-5}{\psframebox{\dfva{c_1/c_1c_2}{\pt{z}{x}}}}
\putnode[l]{w}{r2}{0}{5}{\psframebox{\dfva{c_1c_2/c_1c_2c_2}{\pt{z}{x}}}}
\putnode[l]{w}{n6}{0}{-5}{\psframebox{\dfva{c_1/c_1c_2}{\emptyset}}}
\putnode[l]{w}{ep}{3}{-6}{\psframebox{\dfva{c_1}{\pt{z}{?}}}}
\putnode[l]{w}{ep}{3}{-10}{\psframebox{\dfva{c_1c_2/c_1c_2c_2}{\pt{z}{x}}}}
\end{pspicture}
\end{tabular}
\caption{First round of interprocedural liveness and points-to analysis on our example program. 
Liveness and points-to information 
is subscripted with $L$ and $A$ respectively.
To avoid proliferation of symbols, set of live variables
are represented as strings and `\{' and `\}' are dropped. 
Multiple call strings with the same data flow value are separated by a `/'.
}
\label{fig:ipa.pta.result.1}
\end{figure}

\subsubsection{Matching Contexts for Liveness and Points-to Analysis.}

Since points-to information should be restricted to live ranges,
it is propagated along the call strings constructed
during liveness analysis. However in the presence of recursion, we may
need additional call strings for which liveness information may not
be available. We explain below how this is handled.

Let \acs denote an acyclic call string (i.e.\ a call string for
an interprocedural control flow path with no unfinished recursive
calls). Let \ccs{i} denote a cyclic call string which corresponds to an
interprocedural control flow path with unfinished recursive calls;
$\alpha$ denotes an acyclic sequence of call sites corresponding to
unfinished recursive calls and $i$ denotes the depth of recursion in the
path. Then:
\begin{itemize}
\item The partitioning information for every \acs is available because either
      \sdfv{\acs}{x} has reached node $n$ in procedure $p$ or \acs has
      been represented by some other call string.
\item Assume that the data flow values of \ccs{i} are different for 
      \text{$i \leq k$} for some \text{$k \geq 0$} and the data flow values
      of \ccs{k} and \ccs{k+j}, \text{$j \geq 1$} are identical. Then
      the partitioning information is available for only \ccs{k} and
      \ccs{k+1} because the call strings \ccs{k+j}, \text{$j > 1$} are
      not constructed.
\end{itemize}

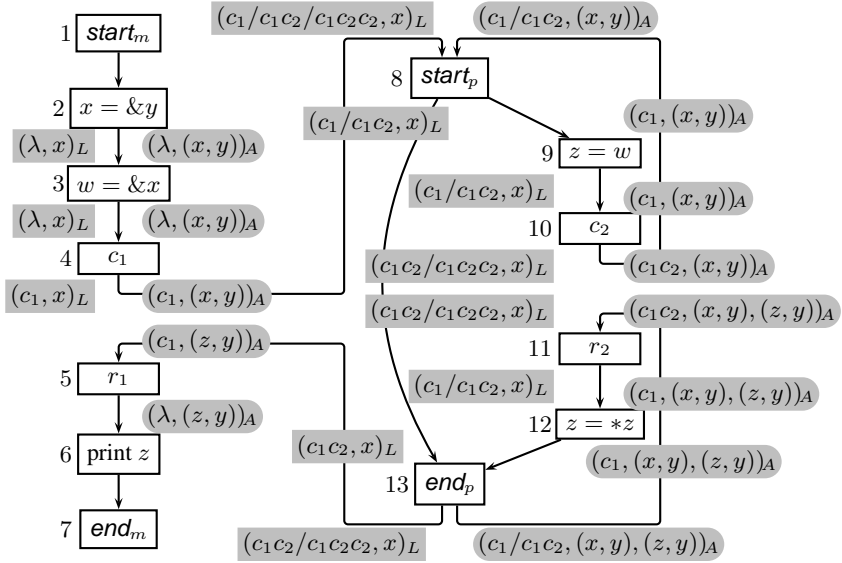
\begin{figure}[!t]
\renewcommand{\pt}[2]{\text{$(#1,#2)$}}
\centering
\begin{tabular}{c}
\begin{pspicture}(0,0)(116,74)
\putnode{entry}{origin}{20}{70}{\psframebox{\Start{m}}}
\putnode{w}{entry}{-7}{0}{$1$}
\putnode{n1}{entry}{0}{-10}{\psframebox{$x = \&y$}}
\putnode{w}{n1}{-8}{0}{$2$}
\putnode{n2}{n1}{0}{-10}{\psframebox{$w = \&x$}}
\putnode{w}{n2}{-8}{0}{$3$}

\putnode{c1}{n2}{0}{-10}{\psframebox{$\;\;\;c_1\;\;\;$}}
\putnode{w}{c1}{-7}{0}{$4$}
\putnode{r1}{c1}{0}{-16}{\psframebox{$\;\;\;r_1\;\;\;$}}
\putnode{w}{r1}{-7}{0}{$5$}
\putnode{n4}{r1}{0}{-10}{\psframebox{print {$z$}}}
\putnode{w}{n4}{-7}{0}{$6$}
\putnode{exit}{n4}{0}{-10}{\psframebox{\End{m}}}
\putnode{w}{exit}{-7}{0}{$7$}
\putnode{sp}{entry}{44}{-6}{\psframebox{\Start{p}}}
\putnode{w}{sp}{-7}{0}{$8$}
\putnode{n5}{sp}{20}{-10}{\psframebox{$z=w$}}
\putnode{w}{n5}{-7}{0}{$9$}
\putnode{c2}{n5}{0}{-10}{\psframebox{$\;\;\;c_2\;\;\;$}}
\putnode{w}{c2}{-8}{0}{$10$}
\putnode{r2}{c2}{0}{-16}{\psframebox{$\;\;\;r_2\;\;\;$}}
\putnode{w}{r2}{-8}{0}{$11$}
\putnode{n6}{r2}{0}{-10}{\psframebox{$z = *z$}}
\putnode{w}{n6}{-8}{0}{$12$}
\putnode{ep}{n6}{-20}{-8}{\psframebox{\End{p}}}
\putnode{w}{ep}{-7}{0}{$13$}
\ncline{->}{entry}{n1}
\ncline{->}{n1}{n2}
\ncline{->}{n2}{n3}
\ncline{->}{n2}{c1}
\ncline{->}{r1}{n4}
\ncline{->}{n4}{exit}
\ncline{->}{sp}{n5}
\ncline{->}{n5}{c2}
\ncline{->}{r2}{n6}
\ncline{->}{n6}{ep}
\nccurve[angleA=240,angleB=120,ncurv=.8]{->}{sp}{ep}
\ncloop[angleA=270,angleB=90,loopsize=-8,linearc=.5,offsetB=1,arm=2.5]{->}{c2}{sp}
\ncloop[angleA=270,angleB=90,loopsize=-27,linearc=.5,offsetA=1,arm=2.5]{->}{ep}{r2}
\ncloop[angleA=270,angleB=90,loopsize=-30,linearc=.5,offsetB=-1,arm=2.5]{->}{c1}{sp}
\ncloop[angleA=270,angleB=90,loopsize=13,linearc=.5,offsetA=-1,arm=2.5]{->}{ep}{r1}
\psset{framesep=.5,linestyle=none,fillstyle=solid,fillcolor=lightgray}
\putnode[r]{w}{n6}{-6}{5}{\psframebox{\dfvl{c_1/c_1c_2}{x}}}
\putnode[r]{w}{r2}{-6}{5}{\psframebox{\dfvl{c_1c_2/c_1c_2c_2}{x}}}
\putnode[r]{w}{ep}{-6}{5}{\psframebox{\dfvl{c_1c_2}{x}}}
\putnode[r]{w}{ep}{-3}{-8}{\psframebox{\dfvl{c_1c_2/c_1c_2c_2}{x}}}
%
\putnode[r]{w}{sp}{0}{-6}{\psframebox{\dfvl{c_1/c_1c_2}{x}}}
\putnode[r]{w}{sp}{-2}{8}{\psframebox{\dfvl{c_1/c_1c_2/c_1c_2c_2}{x}}}
\putnode[r]{w}{c2}{-6}{-5}{\psframebox{\dfvl{c_1c_2/c_1c_2c_2}{x}}}
\putnode[r]{w}{n5}{-6}{-5}{\psframebox{\dfvl{c_1/c_1c_2}{x}}}
\putnode[r]{w}{n2}{-3}{5}{\psframebox{\dfvl{\lambda}{x}}}
\putnode[r]{w}{c1}{-3}{5}{\psframebox{\dfvl{\lambda}{x}}}
\putnode[r]{w}{c1}{-3}{-5}{\psframebox{\dfvl{c_1}{x}}}
\psset{framearc=2}
\putnode[l]{w}{n2}{3}{5}{\psframebox{\dfva{\lambda}{\pt{x}{y}}}}
\putnode[l]{w}{c1}{3}{5}{\psframebox{\dfva{\lambda}{\pt{x}{y}}}}
\putnode[l]{w}{c1}{3}{-5}{\psframebox{\dfva{c_1}{\pt{x}{y}}}}
\putnode[l]{w}{sp}{3}{8}{\psframebox{\dfva{c_1/c_1c_2}{\pt{x}{y}}}}
\putnode[l]{w}{ep}{3}{-8}{\psframebox{\dfva{c_1/c_1c_2}{\pt{x}{y},\pt{z}{y}}}}
\putnode[l]{w}{n5}{3}{5}{\psframebox{\dfva{c_1}{\pt{x}{y}}}}
\putnode[l]{w}{n5}{3}{-6}{\psframebox{\dfva{c_1}{\pt{x}{y}}}}
\putnode[l]{w}{c2}{3}{-5}{\psframebox{\dfva{c_1c_2}{\pt{x}{y}}}}
\putnode[l]{w}{r2}{3}{5}{\psframebox{\dfva{c_1c_2}{\pt{x}{y},\pt{z}{y}}}}
\putnode[l]{w}{r2}{3}{-6}{\psframebox{\dfva{c_1}{\pt{x}{y},\pt{z}{y}}}}
\putnode[l]{w}{n6}{-2}{-5}{\psframebox{\dfva{c_1}{\pt{x}{y},\pt{z}{y}}}}
\putnode[l]{w}{r1}{3}{5}{\psframebox{\dfva{c_1}{\pt{z}{y}}}}
\putnode[l]{w}{n4}{3}{5}{\psframebox{\dfva{\lambda}{\pt{z}{y}}}}
\end{pspicture}
\end{tabular}
\caption{Second round of liveness and points-to analysis to compute 
         dereferencing liveness and the resulting points-to information. 
         Only the additional information is shown.}
\label{fig:ipa.pta.result.2}
\end{figure}

Consider a call string $\sigma'$ reaching node $n$ during points-to
analysis. From the above observations about partitioning it is clear that,
if $\sigma'$ is an acyclic
call string then its partitioning information and hence its liveness
information is available. If $\sigma'$ is a cyclic call string, its
value may not be available if it happens to be \ccs{k+j}, \text{$j >
1$}. However, it is sufficient to locate the longest prefix of \ccs{k+j}
and use its liveness information.
This is illustrated below in our motivating example.

\subsubsection{Motivating Example Revisited.}

{

For brevity, let $I_n$ and $O_n$ denote the entry and exit of node $n$.
In the first round of liveness (Fig.~\ref{fig:ipa.pta.result.1}),
       $z$ becomes live at $I_6$ as \dfvl{\lambda}{z}, reaches 13, 12,
	and 11 as \dfvl{c_1}{z}, becomes \dfvl{c_1c_2}{z} at $I_{11}$, reaches $O_{13}$
	and gets represented by \dfvl{c_1}{z}. Hence 
	\dfvl{c_1c_2}{z} is not propagated within the body of $p$.
        \dfvl{c_1c_2}{z} is regenerated at $I_8$, becomes \dfvl{c_1}{z} at $I_{10}$,
	becomes \dfvl{c_1}{w} at $I_9$. At $O_8$, it combines with \dfvl{c_1}{z}
	propagated from $I_{13}$ and becomes \dfvl{c_1}{w\;z}. Thus $c_1c_2$ is
	regenerated as \dfvl{c_1c_2}{w\;z} at $I_8$.
        \dfvl{c_1}{w\;z} reaches 4 and becomes \dfvl{\lambda}{w\;z}. 

In the first round of points-to analysis (Fig.~\ref{fig:ipa.pta.result.1}),
	since $z$ is live $I_1$, \text{$\boundary = \dfva{\lambda}{\pt{z}{?}}$}.
	\dfva{\lambda}{\pt{w}{x}} is generated at $O_3$. Thus 
	\dfva{c_1}{\pt{w}{x}, \pt{z}{?}} reaches $I_8$. 
        This becomes
	\dfva{c_1}{\pt{w}{x}, \pt{z}{x}} at $O_9$ and reaches as 
	\dfva{c_1c_2}{\pt{w}{x}, \pt{z}{x}} at $I_8$. 
	Since $z$ is not live at $I_9$, \dfva{c_1c_2}{\pt{w}{x}} is propagated to $I_9$ 
	This causes
	\dfva{c_1c_2c_2}{\pt{w}{x}, \pt{z}{x}} to be generated $O_{10}$ which reaches
	$I_9$ and is represented by  \dfva{c_1c_2}{\pt{w}{x}, \pt{z}{x}}. 
	This is then regenerated as  \dfva{c_1c_2c_2}{\pt{z}{x}} at $O_{13}$ because only
	$z$ is live at $O_{13}$. Note that we do not have the liveness information
	along $c_1c_2c_2$ but we know that it must be the same as the liveness information
	along $c_1c_2$.
	We get \dfva{c_1c_2}{\pt{z}{x}} and \dfva{c_1}{\pt{z}{x}} at $O_{11}$. Since we have
	no points-to information for $x$, we get  
	\dfva{c_1c_2}{\emptyset} and \dfva{c_1}{\emptyset} at $O_{12}$.

The second round of liveness and points-to analysis is presented in 
	Fig.~\ref{fig:ipa.pta.result.2}. We leave it for the reader to verify that
	$x$ becomes live due to \text{$z=*z$} in 12, reaches 2 and causes \dfva{\lambda}{x}{y}
	to be generated. As a consequence, we get \pt{z}{y} in 12. 

This result corresponds to the observations in Section~\ref{sec:intro}.
	Note that \pt{z}{x}
	cannot reach 6 along any interprocedurally valid path. However, 
        the method of~\cite{Emami.M.Ghiya.R.Hendren.LJ:1994:Context-sensitive-interprocedural-points-to}
        which is considered most precise flow- and context-sensitive method, computes \pt{z}{x}
	at 6.

}

\section{Related Work}
\label{sec:related.work}

The benefits of flow- and context-sensitivity have been found to 
vary from marginal to large in the literature~\cite{%
Ruf.E:1995:Context-Insensitive-Alias-Analysis,%
Lhotak.O.Hendren.LJ:2006:Context-sensitive-points-to-analysis,%
Shapiro.M.Horwitz.S:1997:Effects-of-Precision,%
Hind.M.Pioli.A:1998:Assessing-Effects-of}. 
It has also been observed that  an increase in precision could 
increase efficiency. However, studies have been inconclusive by and large
and a large number of investigations relax
flow- or context-sensitivity (or both) in their pursuit of efficiency in
pointer analysis.  Our premise is that the use of liveness enhances
the effectiveness of flow- and context-sensitivity significantly.
A flow-insensitive approach cannot benefit from liveness.
The use of liveness in context-insensitive approaches
has not been investigated. 

We focus on approaches that are both flow- and context-sensitive. 
A memoisation-based functional approach observes that
the number of possible pointer patterns
that reach a procedure are small and hence it is beneficial to use
partial transfer
functions~\cite{Wilson.RP.Lam.MS:1995:Efficient-Context-Sensitive-Pointer} 
instead of the usual full transfer functions.
An alternative functional approach creates full transfer
functions but contains the complexity of computing transfer functions
by making them sensitive to the ``level'' of a pointer (i.e. the
possible depth of its
indirection)~\cite{Yu.H.Xue.J.Huo.W.Feng.X.ea:2010:Level-by-level}.
Transfer functions for a given level are defined in terms of
lower-level transfer functions. 
The invocation-graph-based approach unfolds a call graph in terms
of call chains~\cite{Emami.M.Ghiya.R.Hendren.LJ:1994:Context-sensitive-interprocedural-points-to}.
Our work is inspired by this approach but we have incorporated strong liveness
and manage contexts very differently.
Finally, a radically different approach
proceeds in the opposite direction and begins with flow- and
context-insensitive information which is refined systematically
in cascaded steps to restrict it to flow- and context-sensitive
information~\cite{Kahlon.V:2008:Bootstrapping-technique-for}.

The above approaches summarise points-to information in recursive contexts
using fixed-point iteration.
This merges the information across different levels
of nesting and all recursive calls receive the same summarised
information.
The call-strings approach maintains distinct data-flow values for
each nesting depth of recursion. 
The partial-transfer-function-based approach~\cite{Wilson.RP.Lam.MS:1995:Efficient-Context-Sensitive-Pointer}
is slightly more precise than the invocation-graph-based 
approach~\cite{Emami.M.Ghiya.R.Hendren.LJ:1994:Context-sensitive-interprocedural-points-to}
because it distinguishes the outer call to a recursive procedure
from the calls inside the recursion. For example, in our motivating
example, \pt{z}{x} holds only in the recursive calls of $p$.
When recursion unwinds fully, $z$ does not point to $x$. Our approach
discovers this correctly but~\cite{Emami.M.Ghiya.R.Hendren.LJ:1994:Context-sensitive-interprocedural-points-to}
cannot do so. Fig. 9.6 (page 305) in~\cite{Khedker.U.Sanyal.A.Karkare.B:2009:Data-Flow-Analysis}
contains an example for which the methods 
in~\cite{Wilson.RP.Lam.MS:1995:Efficient-Context-Sensitive-Pointer,%
Emami.M.Ghiya.R.Hendren.LJ:1994:Context-sensitive-interprocedural-points-to}
compute imprecise results.

GCC uses a context-insensitive analysis which acquires limited
flow sensitivity due to the effect of SSA representation---a half-way house.
However, SSA form does not apply to 
pointers directly and interleaved SSA construction and pointer analysis
are required~\cite{Hasti.R.Horwitz.S:1998:Using-static-single} which is not
done in GCC\@.
Appendix~\ref{app:pt.counter.examples} shows by example that
the points-to information in GCC is effectively flow-insensitive.


\section{Implementation and Empirical Measurements}
\label{sec:measurements}

We have implemented interprocedural lazy points-to analysis in GCC 4.6.0.
It requires the command line switches 
{\tt\bfseries -flto -flto-partition=none -flipta}
to invoke GCC's Link Time Optimisation (LTO), pass on the control
flow and call graphs, and finally perform lazy points-to analysis on
the constructed supergraph. This implementation 
is available for download.\footnote{%
\htmladdnormallink%
{\tt\bfseries http://www.cse.iitb.ac.in/grc/index.php?page=lipta}%
{http://www.cse.iitb.ac.in/grc/index.php?page=lipta}.}


We have executed our implementation on SPEC CPU2006 Integer 
benchmarks
as well as 
some programs from SPEC2000 benchmarks on a machine with 16 GB RAM
running 8 processors (64-bit intel i7-960 CPU at 3.20GHz).
The results of measurements are presented in
Fig.~\ref{fig:measurements}. We compare three implementations:
{\em lazy points-to analysis (lpta)},
{\em simple points-to analysis (spta)} and
{\em GCC's points-to analysis (gpta)}.
The only difference between lpta and spta
is that lpta uses liveness whereas spta does not---both are flow- and
context-sensitive and use call strings with value-based termination.
gpta is flow- and context-insensitive (see Section~\ref{sec:related.work} for
more details about GCC's points-to analysis). All three methods use the
same approach of handling arrays, heap locations, pointer arithmetic,
function pointers, and field sensitivity.

\begin{figure}[t]
\begin{center}
\begin{tabular}{| l | r | r  		
		| r | r | r | r | r  	
		| r | r | r  		
		| r			
		|	}
\hline
\multicolumn{11}{|c|}{
	lpta = Lazy PTA,
	spta = Simple PTA,
	gpta = GCC's PTA}
	\\ \hline
\multirow{3}{*}{Program} 
	& \multicolumn{1}{c|}{\multirow{3}{*}{\rotatebox{90}{kLoC}}}
	& \multicolumn{1}{c|}{\multirow{3}{*}{\begin{tabular}{@{}l@{}}
			Call \\
			Sites
			\end{tabular}}}
	& \multicolumn{4}{c}{Time in milliseconds} 
	& \multicolumn{3}{|c|}{Points-to pairs} 
	& \multicolumn{1}{c|}{\multirow{3}{*}{\rotatebox{90}{\!\!Max\#cs}}}
	\\ \cline{4-10}
  & &
	& \multicolumn{2}{c|}{lpta}
	& \multicolumn{1}{c|}{\multirow{2}{*}{spta}}
	& \multicolumn{1}{c|}{\multirow{2}{*}{gpta}}
	& \multicolumn{1}{c|}{\multirow{2}{*}{lpta}}
	& \multicolumn{1}{c|}{\multirow{2}{*}{spta}}
	& \multicolumn{1}{c|}{\multirow{2}{*}{gpta}}
	&
		\\ \cline{4-5}
  & & 
	& \multicolumn{1}{c|}{liveness}
	& \multicolumn{1}{c|}{pta}
	& & & & & &
	\\ \hline\hline
lbm 
	& 0.9 
	& 33 
	& 0.55 
	& 0.52 
	& 1.9 
	& 5.2 
	& 12 
	& 507 
	& 1911  
	& 4
	\\ \hline
mcf 
	& 1.6 
	& 29 
	& 1.04 
	& 0.62 
	& 9.5 
	& 3.4 
	& 41 
	& 367 
	& 2159  
	& 4
	\\ \hline 
libquantum
	& 2.6 
	& 258 
	& 2.0 
	& 1.8 
	& 5.6 
	& 4.8 
	& 49 
	& 119 
	& 2701  
	& 55
	\\ \hline
bzip2 
	& 3.7 
	& 233 
	& 4.5 
	& 4.8 
	& 28.1 
	& 30.2 
	& 60 
	& 210 
	& 8.8$\times 10^4$  \rule{0em}{1em}
	& 70
	\\ \hline
parser 
	& 7.7 
	& 1123 
	& 1.2$\times 10^3$ 
	& 145.6 
	& 4.3$\times 10^5$ 
	& 422.12 
	& 531 
	& 4196 
	& 1.9$\times 10^4$  \rule{0em}{1em}
	& 4619
	\\ \hline
sjeng 
	& 10.5 
	& 678 
	& 858.2 
	& 99.0 
	&3.2$\times 10^4$ 
	& 38.1 
	& 267 
	& 818 
	& 1.1$\times 10^4$  \rule{0em}{1em}
	& 4649
	\\ \hline
hmmer
	& 20.6 
	& 1292 
	& 90.0 
	& 62.9 
	& 2.9$\times 10^5$ 
	& 246.3 
	& 232 
	& 5805 
	& 1.9$\times 10^6$  \rule{0em}{1.1em}
	& 554
	\\ \hline
gap 
	&  35.6 
	&  5312 
	&  4.6$\times 10^4$ 
	&  1.3$\times 10^3$ 
	&  1.0$\times 10^5$ 
	&  1.7$\times 10^4$ 
	&  421 
	&  1271 
	&  2.5$\times 10^9$ \rule{0em}{1.1em}
	& 1203
	\\ \hline
h264ref 
	& 36.0 
	& 1992 
	& 2.2$\times 10^5$ 
	& 2.0$\times 10^5$ 
	& ? 
	& 4.3$\times 10^3$ 
	& 1683 
	& ? 
	& 1.6$\times 10^7$ \rule{0em}{1em}
	& 46660
	\\ \hline
\end{tabular}
\end{center}
\caption{Empirical measurements.  A `?' indicates
that the analysis ran out of memory. Max\#cs denotes maximum number of call strings at any program point
for lpta.}
\label{fig:measurements}
\end{figure}

Both lpta and spta are naive implementations that use linked lists and
linear searches within them.
The main goal of these implementations
was to find out whether liveness increases the precision of points-to
information. Our measurements confirm this hypothesis beyond doubt.
Surprisingly, the time measurements exceeded our expectations because we
had not designed these implementation for time/space efficiency or
scalability. We were able to run our implementations on programs of
around 30kLoC but not on the larger programs. It is evident from the
measurements that:
\begin{itemize}
\item Lazy computation of points-to pairs reduces the number of 
      points-to pairs dramatically. Although we
      could observe this for programs of approximately 30kLoC, we have no 
      reason to believe that the situation would be different for larger programs.
\item Lazy computation and sparse propagation of points-to pairs reduces 
      execution time too and lpta out-performs gpta for most programs smaller than 30kLoC.
      That a flow- and context-sensitive analysis could be faster than
      flow- and context-insensitive analysis comes as a surprise to
      us. lpta shows that the actual data that we can gainfully use is
      much smaller than what is generally thought to be.
\item A reduction in the number of data flow values enhances the effectiveness of
      value based termination of call strings and in most cases the
      number of contexts required for precise analysis is not
      exponentially large.  Further, 
      the maximum length of any call string never exceeded two digits.
\end{itemize}

The hypothesis that our implementation suffers because of
linear search in linked lists
was confirmed by an accidental discovery: in order to eliminate
duplicate pairs in gpta, we used our data structure and function
from lpta that adds points-to pairs in a linked list and maintains
a unique entry for each pair in the list. With this addition, gpta
executed for well over an hour on the hmmer program
whereas originally gpta needed 246.3 milliseconds only!
Since lpta uses linked lists to represent sets, it has to maintain
uniqueness at each stage and this seems to be the primary reason why we
could not execute it on the larger programs: gobmk, perlbench, and gcc.

Eager liveness computation to reduce points-to analysis work could also
be a source of inefficiency: a new round of liveness is invoked when
a new points-to pair for $y$ is discovered for \text{$x = *y$} putting on hold
the points-to analysis. This explains the unusually large time spent in liveness
analysis  compared to points-to analysis
for programs parser and sjeng. The number of 
rounds of analysis required for these programs was much higher than in other programs
of comparable size.

Our implementation can be improved many ways.
\begin{itemize}
\item We can use efficient data structures (vectors or hash tables)
      supported by GCC\@. Alternatively, we can use
      BDDs to efficiently maintain sets of data flow values.
\item We can experiment with less eager strategies of invoking liveness analysis.
\item The LTO framework could be modified to load CFGs on demand.
      Currently,
      LTO gives one large program with all CFGs or just a call
      graph without CFGs.
      This results in a very large supergraph in memory---affecting
      locality (cache misses) partly explaining the 30kLoC threshold.
\item Our implementation performs full computations of liveness
      and points-to analysis.
      Revisiting a statement typically causes only
      a small additional amount of information to be generated.
      We posit significant savings by exploiting this third
      dimension of laziness: compute information incrementally on revisits.
\end{itemize}

\noindent
Apart from improving the implementation, another route to scalability 
lies in the observation that 30kLoC seems to be a cross-over point: 
If we can preprocess programs to identify chunks of around 30kLoC which
are very loosely coupled as far as pointer usage is concerned,
we can expect this method to scale to much larger programs.

\section{Conclusions and Future Work}
\label{sec:conclusions}

We have described a data-flow analysis which jointly calculates
points-to and liveness information.
It does this in a flow- and context-sensitive way, using recent developments
of the ``call strings'' approach.
One novel aspect to our approach is that it is effectively bi-directional
(such analysis seem relatively rarely exploited).

Initial results from our naive prototype implementation were impressive:
unsurprisingly our analysis produced much more precise results, but
by an order of magnitude (in terms of the size of the calculated
points-to information).  The reduction of this size allowed our
naive implementation also to run faster than GCC's points-to
analysis at least for programs up to 30kLoC. This is significant because
GCC's analysis compromises both on flow and context sensitivity.
This confirms our belief that separating relevant information from
irrelevant information can have significant benefits and is a promising
direction for further investigations.

We would like to take our work further by exploring the following:
\begin{itemize}
\item Improving our implementation: e.g.\ using
efficient data structures such as vectors or hash tables,
or perhaps BDDs.
Improving the interface to GCC's LTO framework by allowing the call graph
to be loaded as a single unit, but then loading
individual CFGs on demand so as not to keep the whole-program
supergraph in memory at one time.
\item Exploring the reasons for the 30kLoC speed threshold;
while interprocedural analyses are very likely to be super-linear
in terms of the number of procedures, perhaps there are ways in practice to
partition most bigger programs (around loosely-coupled boundaries)
without significant loss of precision.
\item Currently our use of incremental computation is solely to avoid
computing useless and imprecise data-flow information.  However, we note
that data-flow information often only slightly changes when revisiting a node
compared to the information produced by the first iteration.
We plan to explore incremental formulations of our lazy points-to analysis.
\end{itemize}

\section*{Acknowledgements}
Empirical measurements were carried out by Prachee Yogi and Aboli Aradhye. Prachee also implemented a prototype of intraprocedural
analysis in Prolog.  Ashwin Paranjape acted as a sounding board for our initial ideas.

\bibliography{lazy-pta-abbrv.bib}

\appendix
\setcounter{lemma}{0}
\setcounter{theorem}{0}
\setcounter{corollary}{0}

\section{Proofs of Lemmas and Theorems}
\label{app:proofs}

\begin{theorem}
Define the liveness flow function \text{$f_L : \lvE\times\aptE
\mapsto \lvE$} (Equation~\ref{eq:lin.exhaustive}) and the may-points-to
flow function \text{$f_A : \lvE\times\aptE \mapsto \aptE$}
(Equation~\ref{eq:aout.exhaustive}) as:
\begin{align*}
f_L (l,a) & = (l - \Kill\,(a)) \cup \Ref\,(l,a)
	\\
f_A (l,a) & = \bigrestrict{\left(a - (\Kill\,(a)\times \var)\right) 
	\cup \left(\Def\,(a)\times\Pointee\,(a)\right)}{l}
\end{align*}
Then, $f_L$ and $f_A$ are monotonic.
\end{theorem}
\begin{proof}
Monotonicity of  $f_L$ can be proved by showing that 
\text{$\forall (l_1, a_1), (l_2, a_2) \in \lvE\times\aptE$},
\begin{align}
(l_1,a_1) \sqsubseteq (l_2,a_2)  
	& \Rightarrow   \left(l_1 - \Kill\,(a_1)\right) \supseteq \left(l_2 - \Kill\,(a_2)\right)
	\label{po.lv.mo.1}
	\\
(l_1,a_1) \sqsubseteq (l_2,a_2)  
	& \Rightarrow   \Ref\,(l_1,a_1) \supseteq \Ref\,(l_2,a_2)
	\label{po.lv.mo.2}
\end{align}
(\ref{po.lv.mo.1}) follows from~(\ref{eq:monotonicity.5}) while
(\ref{po.lv.mo.2}) follows from~(\ref{eq:monotonicity.1}).
Monotonicity of  $f_A$ can be proved by showing that 
\text{$\forall (l_1, a_1), (l_2, a_2) \in \lvE\times\aptE$},
\begin{align}
(l_1,a_1) \sqsubseteq (l_2,a_2)  
	& \Rightarrow   \left(a_2 - (\Kill\,(a_1) \times \var )\right) 
		\supseteq \left(a_1 - \left(\Kill\,(a_2) \times \var\right)\right)
	\label{po.pt.mo.1}
	\\
(l_1,a_1) \sqsubseteq (l_2,a_2)  
	& \Rightarrow   \left(\Def\,(a_1)\times\Pointee\,(a_1)\right) \supseteq 
		\left(\Def\,(a_2)\times\Pointee\,(a_2)\right)
	\label{po.pt.mo.2}
\end{align}
(\ref{po.pt.mo.1}) follows from~(\ref{eq:monotonicity.5}) while
(\ref{po.pt.mo.2}) follows from~(\ref{eq:monotonicity.3}) and~(\ref{eq:monotonicity.4}). \qed
\end{proof}

\begin{theorem}
Every propagation path for a points-to pair \pt{x}{y} is a suffix of some 
liveness path for $x$.
\end{theorem}
\begin{proof}
Consider an arbitrary propagation path $\rho_a$ for \pt{x}{y}. Since
\text{$\text{L2} \Leftrightarrow \text{A2}$}, \text{$\text{A3}
\Rightarrow \text{L3}$}, and \text{$\text{A4} \Rightarrow \text{L4}$},
it is easy to see that every statement $n$ along $\rho_a$ must also be
part of a liveness path for $x$. Let this liveness path be $\rho_l$.
Then the proof obligation reduces to showing that the last statement of
$\rho_a$ must also be the last statement of $\rho_l$. In other words, we
need to show that
\begin{enumerate}
\item[C1.] $\rho_a$ does not end somewhere in the middle of $\rho_l$, and 
\item[C2.] $\rho_a$ does not extend beyond $\rho_l$.
\end{enumerate}
We prove these by contradiction. For case (C1), assume that the last
statement of $\rho_a$ appears somewhere in the middle of $\rho_l$ on
position $j$. Consider statements $s_j$ and $s_{j+1}$ in $\rho_l$
such that $s_j$ also appears in $\rho_a$. From (L3) and (L4), in
\text{$x \in\lin_{s_{j}}$}, \text{$x \in\lout_{s_{j}}$}, and \text{$x
\in\lin_{s_{j+1}}$}. Also, $x$ is neither in $\Kill_{s_{j}}$ nor
$\Kill_{s_{j+1}}$ from (L3). Further \text{$\pt{x}{y} \in \ain_{s_j}$}
from (A2).
\begin{eqnarray*}
\pt{x}{y} \in \ain_{s_j} \wedge x \not\in \Kill_{s_j} \wedge x \in \lout_{s_j}
	& \Rightarrow & \pt{x}{y} \in \aout_{s_j}
	\\
	\pt{x}{y} \in \aout_{s_j} \wedge x \in \lin_{s_{j+1}}
	& \Rightarrow & \pt{x}{y} \in \ain_{s_{j+1}}
\end{eqnarray*} 
Thus (A3) is satisfied for $s_{j+1}$ also. Hence $\rho_a$ is not        
maximal and can be extended to include $s_{j+1}$. This leads to         
contradiction.                                                          

For case (C2) assume that the last statement of $\rho_l$ appears
somewhere in the middle of $\rho_a$ on position $j$. Consider
statements $s_j$ and $s_{j+1}$ in $\rho_a$ such that $s_j$ also
appears in $\rho_l$. Then by conditions (A3) and (A4), \text{$x
\not\in \Kill_{s_j}$}, \text{$x \in \lout_{s_j}$}, and \text{$x \in
\lin_{s_{j+1}}$}. Thus $\rho_l$ is not maximal and can be extended to
include $s_{j+1}$. This leads to contradiction. \qed
\end{proof}

\begin{lemma}
\label{lemma.kill.def}
$\left((\Kill_n \neq \emptyset ) \wedge (\Kill_n \neq \pointer)\right) 
       \Rightarrow \left(\Kill_n = \Def_n\right).
$
\end{lemma}
\begin{proof}
The lemma trivially holds for all statements other than indirect
assignment \text{$*x = y$}. For the latter,
\begin{align*}
\left((\Kill_n \neq \emptyset ) \wedge (\Kill_n \neq \pointer)\right) 
	& \Rightarrow 
		\left(\left(\lrestrict{\ain_n}{x}\right) \neq \emptyset\right)
	\wedge \left(\left(\lrestrict{\ain_n}{x}\right) \neq \{\pt{x}{?}\}\right)
	\\
	& \Rightarrow
		\left(\ain_n = \{ \pt{x}{z} \}\right) \wedge (z \in \pointer)
\end{align*}
Hence \text{$\Kill_n = \Def_n = \{z\}$}.
\qed
\end{proof}

\begin{theorem}
\label{theorem.live.pt}
If \text{$x\in \pointer$} holds the address of \text{$z \in (\var - \{?\})$}
along some execution path reaching node $n$, then \text{$x \in \Ref_n \Rightarrow 
\pt{x}{z} \in \ain_n$}.
\end{theorem}
\begin{proof}
Let the execution path reaching node $n$ be denoted by \text{$\rho
\equiv s_0, s_1, \ldots, s_k$} where \text{$s_0 = \Start{p}$} and
\text{$s_k = n$}. We prove the theorem by induction on path length
$k$. The basis is $k=2$ where $s_1$ assigns the address of $z$ to $x$
and $s_2$ uses it.\footnote{When statement $n$ uses $*x$, the minimum
length should be $k=3$ so that the pointee of pointee of $x$ is also
defined but this is not relevant at the moment.} Since \text{$x \in
\lin_{s_2}$}, the sequence $s_1,s_2$ is trivially both a liveness
path as well as points-to propagation path. Thus, \text{$x \in \Ref_n
\Rightarrow \pt{x}{z} \in \ain_n$}.

Assume that the inductive hypothesis holds for $k=i$. Consider the case
when $k=i+1$. Note that \text{$x \in \lout_{s_i}$}. Statement $s_i$
could influence $x$ in the following ways:
\begin{itemize}
\item \text{$x \not\in \Kill_{s_i}$}. Assume that the last node in
      path $\rho$ in which $x$ is assigned a value is $s_m$, \text{$m
      < i$}. Statement $s_m$ either directly assigns $\&z$ to $x$, or
      does so through some variables in $\Ref_{s_m}$. By inductive
      hypothesis, the pointees of every variable in $\Ref_{s_m}$ have
      been discovered in $\ain_{s_m}$. Thus points-to analysis would
      discover that \text{$\pt{x}{z} \in \aout_{s_m}$}. The suffix
      of $\rho$ from ${s_m}$ to ${s_{i+1}}$ is both a liveness path
      for $x$ and points-to propagation path for \pt{x}{z}. Hence
      \text{$\pt{x}{z} \in \ain_n$}.
\item \text{$x \in \Kill_{s_i}$}. In this case, $\Kill_{s_i}$ could be
      \pointer if statement ${s_i}$ is an indirect assignment \text{$*w
      = y$}. Since \text{$w\in\lin_{s_i}$}, by inductive hypothesis
      \text{$\pt{w}{u} \in \ain_{s_i}$} such that \text{$u\neq?$}.
      Hence the first condition of~(\ref{eq:must.new.def}) cannot be
      satisfied. Thus this case is ruled out and \text{$\Kill_{s_i}
      \neq \pointer$}. However since \text{$\Kill_{s_i} \neq
      \emptyset$}, \text{$\Kill_{s_i} = \Def_{s_i} = \{x\}$} from
      Lemma~\ref{lemma.kill.def}. By a reasoning similar to that of node
      $s_m$ in the previous case, \text{$\pt{x}{z} \in \aout_{s_i}$}.
      Since \text{$x \in \lin_{s_{i+1}}$}, the sequence $s_i,s_{i+1}$ is
      trivially both a liveness path as well as points-to propagation
      path. Thus, \text{$\pt{x}{z} \in \ain_n$}.
\end{itemize}
Thus the theorem holds because the inductive hypothesis holds for $k=i+1$.
\qed
\end{proof}

\begin{corollary}
\label{cor.def.not.empty}
If all pointer variables are initialised with values of proper types
before they are used, then for every indirect assignment \text{$*x=y$},
\text{$\Def_n \neq \emptyset$}.
\end{corollary}
\begin{proof}
Since \text{$x\in\Ref_n$}, \text{$\exists\pt{x}{z} \in \ain_n$} such that \text{$z\neq?$}
from Theorem~\ref{theorem.live.pt}. Thus $\Def_n$ cannot be $\emptyset$.
\qed
\end{proof}

\section{Flow Insensitivity in GCC's Points-to analysis}
\label{app:pt.counter.examples} 

Consider the following program:
\begin{verbatim}
#include <stdio.h>
int a, b, c, *e;
int main()
{

        if (a == b)
                e = &c;   /* statement n1 */
        else
                e = &b;   /* statement n2 */
        e = &a;           /* statement n3 */
        p();
}

p()
{
        printf ("%d", e);
}
\end{verbatim}

In a flow sensitive analysis the points-to set of $e$ will not contain $a, b, c$ at the same time. 
There should be four different points-to sets associated with $e$: After {\tt n1} and {\tt n2},
it should be \text{$\{c\}$} and \text{$\{b\}$} respectively whereas it should be \text{$\{b,c\}$}
before {\tt n3} and \text{$\{a\}$} after it.  However,
GCC computes a single points-to set for $e$ that contains all three of them.
The relevant fragment from GCC's dump is as follows:
\begin{verbatim}
Points-to sets

NULL = { }
ANYTHING = { ANYTHING }
READONLY = { READONLY }
ESCAPED = { READONLY ESCAPED NONLOCAL a b c }
NONLOCAL = { ESCAPED NONLOCAL }
CALLUSED = { }
STOREDANYTHING = { }
INTEGER = { ANYTHING }
e.0_1 = same as e
e = { ESCAPED NONLOCAL a b c }
a.1_1 = { ESCAPED NONLOCAL }
a = same as a.1_1
b.2_2 = { ESCAPED NONLOCAL }
b = same as b.2_2
c = { ESCAPED NONLOCAL }
\end{verbatim}
\end{document}